\documentclass{article}

\usepackage[a4paper]{geometry}
\usepackage{amsmath,amssymb}
\usepackage{parskip}
\usepackage{graphicx}
\usepackage[font=small,labelfont=bf,labelsep=endash]{caption}
\usepackage[colorlinks,linkcolor=black,citecolor=black]{hyperref}
\usepackage{tikz}
\usepackage{multirow}

\DeclareMathOperator*{\argmin}{argmin}

\title{Inverting elastic dislocations using the \\ Weakly-enforced Slip Method}
\author{G.J.\ van Zwieten, E.H.\ van Brummelen, R.F.\ Hanssen}

\usepackage[a4paper]{geometry}

\begin{document}

\maketitle

\tableofcontents

\section{Introduction}

With the advent of satellite based interferometry, or InSAR, routine
measurements of the earth's surface deformation have become available,
providing a wealth of information about subsurface processes~\cite{Biggs_2020}.
One of these processes is tectonic faulting, along with its violent
manifestation, earthquakes. While the dynamics of the quake itself cannot be
measured from space in the way that seismometers do, what can be measured to
great accuracy is the lasting adjustments of static equilibrium to the defect
that results from the relative displacement, or \emph{slip}, of two adjacent
masses along a fault.

Given two measurements of the earth's surface covering the area of an
earthquake, of which one taken just prior, the other just after, it can be
reasonably assumed that the differences can be wholly attributed to the process
of tectonic faulting~\cite{Prescott_1993}. Separation of \emph{co-seismic} and
\emph{post-seismic} signal can be controlled by shortening the satellite
revisit time~\cite{Elliott_2016}. However, to establish the details of the
faulting mechanism that corresponds to the observed co-seismic deformation,
such as the location and orientation of the fault, and the amount, depth and
direction of the slip that occurred, we require an understanding of the physics
bridging the two.

It is generally assumed that on the near-instantaneous time scale of co-seismic
deformation, the earth behaves to a good approximation
elastically~\cite{Segall_2010}. The default model connecting the fault
mechanism and surface observations, therefore, is that of static, elastic
dislocations. While a formal definition of this problem will be presented in
Section~\ref{sec:forward}, sufficing at present is that this type of problem
has been studied for well over a century, during which a range of solution
methods has been devised of which \cite{vanZwieten_2013} provide an overview.

Of the many solution methods available, the most powerful is arguably the
finite element method~\cite{Zienkiewicz_2013}, which is capable of
incorporating all available knowledge of material heterogeneity and surface
topography and thus capturing the system to the greatest detail. Regrettably,
for reasons of computational cost, InSAR analyses are commonly performed on the
other end of the complexity spectrum, in the confines of a homogeneous half
space for which analytical expressions are available --- see, for example
~\cite{Simons_2002, Fialko_2005, Wenbin_2015}. It is with this situation in
mind that \cite{vanZwieten_2014} proposed the Weakly-enforced Slip Method
(WSM), in an attempt to combine the power of the finite element method with the
computational efficiency of analytical expressions.

The efficiency of WSM does come at a price: the displacement field it produces
is continuous, and therefore unable to capture the jump at the fault. In
\cite{vanZwieten_2014} it was however shown that the error thus incurred decays
exponentially with distance to the fault. It was therefore hypothesized that
this drawback is of little consequence in the context of satellite
observations, as most surface measurements are sufficiently far removed from
the dislocation. Only rupturing or near-rupturing faults will cause numerical
errors that significantly interfere with the observables, and even that only
locally. As InSAR data tends to be of low quality in these regions with damage
leading to decorrelation, it is hoped that a reduction of numerical accuracy in
this region will be of little consequence.

It bears repeating that the above considerations are speculative. While
\cite{vanZwieten_2014} proved the mathematical soundness of the WSM, it is in
the present paper that we set out to thoroughly test the utility of the WSM in
the problem setting for which it was devised. To this end a number of synthetic
but otherwise fully representative case studies are presented and analysed
using the WSM, as well as validated against the exact solutions. For the sake
of this validation the scenarios will be restricted to homogeneous half spaces
so that exact solutions are available in the form of analytical expressions,
but it is to be understood that the presented methodology is valid for the
wider class of problems including material heterogeneity and topography.

The central problem that is studied in this paper is the following:

\textbf{(Inverse problem)}
\emph{Given observed displacements of the earth's surface, determine the
fault location and slip distribution that are in best accordance with these
observations, as well as an uncertainly measure for this result.}

To solve this problem we first need to address the conjugate question:

\textbf{(Forward problem)}
\emph{Given a fault location and slip distribution, determine the surface
displacements that we can expect to observe.}

The forward problem is the topic of Section~\ref{sec:forward}, the inverse
problem that of Section~\ref{sec:inverse}. Section~\ref{sec:method} will
define the methodology and introduce case studies, and
Section~\ref{sec:results} will present the results of the comparative study.

\section{Forward Problem: Linear Elastic Dislocation}
\label{sec:forward}

To formally define the dislocation problem that will stand model to the earth's
response to tectonic faulting, we will denote by $\Omega$ the solid domain, by
$d$ its spatial dimension, and by $u: \Omega \rightarrow \mathbb R^d$ the
deformation field, i.e.\ the displacement of the solid compared to its
reference configuration. We assume that it is possible to create a mapping from
the deformation field $u$ to the corresponding state of internal stress
$\sigma$. In particular, we assume that the stress depends linearly on the
deformation gradient, leading to the well known constitutive relation
\begin{equation}
\sigma(u) = C : \nabla u,
\label{eqn:stressstrain}
\end{equation}
where $C: \Omega \rightarrow \mathbb R^{d\times d\times d\times d}$ is the
stiffness tensor representing local material properties. For the medium to be
at rest, Newton's second law states that the divergence of stress must be in
balance with the applied loading. Rather than incorporating the loading
conditions of the earth's gravitational field, we use the linearity of the
stress-strain relation of Equation~\eqref{eqn:stressstrain} to have $u$
represent only the deviations relative to the existing equilibrium. The earth's
gravitational field being constant in time, this means the stress field
\eqref{eqn:stressstrain} is divergence free.

Through the remainder of this document we will use a consistent notation for
the four spaces that form the basis of our mathematical framework. By $\mathcal
X$ we denote the space of local fault plane coordinates $\xi \in \mathbb
R^{d-1}$. By $\mathcal M$ we denote the space of all possible fault geometries
$m: \mathcal X \rightarrow \mathbb R^d$ that position the manifold in physical
space. By $\mathcal B$ we denote the space of slip distributions pulled back to
$\mathcal X$, the slip vector $b(\xi)$ being the jump in the displacement field
when passing from one side of the manifold to the other. By $\mathcal D$ we
denote the space of surface displacements $d: \mathbb R^{d-1} \rightarrow
\mathbb R$ as measured in line of sight to the satellite. With this notation in
place we can reformulate the forward problem as follows:

\textbf{(Forward problem)}
\emph{Determine the surface observations $d \in \mathcal D$ corresponding to a
given manifold $m \in \mathcal M$ and slip distribution $b \in \mathcal B$.}

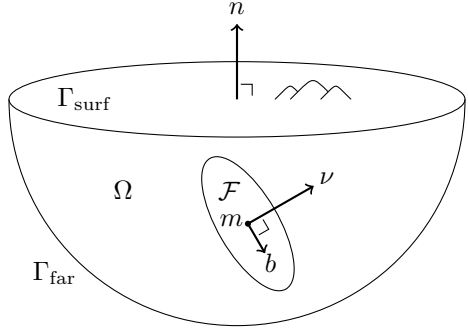
\begin{figure}
\centering
\begin{tikzpicture}
\draw (-3,0) arc (-180:0:3);
\draw (0,0) ellipse (3 and .5);
\draw [thick,->] (0,0) -- (0,1) node[anchor=south] {$n$};
\draw (.05,.2)--(.2,.2)--(.2,.05);
\begin{scope}[rotate=30,shift={(-.7,-1.5)}] 
  \draw (0,0) ellipse (.4 and 1);
  \fill (0,0) circle (.4mm);
  \draw (.05,-.2)--(.2,-.2)--(.2,-.05);
  \draw [thick,<->] (0,-.45)--(0,0)--(1,0);
  \draw (1.2,0) node {$\nu$};
  \draw (0,-.6) node {$b$};
  \draw (-.16,.15) node {$m$};
  \draw (0,.5) node {$\mathcal F$};
\end{scope}
\begin{scope}[shift={(1,0)}] 
  \draw (-.5,0)--(-.4,.1).. controls (-.3,.2)..(-.2,.1);
  \draw (-.3,0)--(-.2,.1).. controls (  0,.3)..( .2,.1);
  \draw ( .1,0)--( .2,.1).. controls ( .3,.2)..( .4,.1)--(.5,0);
\end{scope}
\draw (-2,0) node {$ \Gamma_{\rm surf} $};
\draw (-2.4,-2.3) node {$ \Gamma_{\rm far} $};
\draw (-1.5,-1.2) node {$ \Omega $};
\end{tikzpicture}
\caption{Adapted from~\cite{vanZwieten_2013}, visualization of the forward problem and all entities of Equation~\eqref{eqn:forward}.}
\label{fig:forward}
\end{figure}

We denote the fault plane $\mathcal F$ and the domain boundary $\Gamma$ as
shown in Figure~\ref{fig:forward}. At the fault $\mathcal F = m(\mathcal X)$ we
demand that the displacement field $u$ jumps discontinuously by a distance $b$.
Since this makes the displacement field locally non-differentiable, the stress
is locally not defined, and our general equilibrium condition does not apply.
Instead, Newton's second law transforms into a jump condition for the traction,
stating that the tractions on either side of the fault plane must be in
balance. At the earth's surface $\Gamma_{\rm surf}$ we assume traction-free
conditions, and the remaining domain boundary $\Gamma_{\rm far}$ is assumed to
be sufficiently far away for the relative displacements to be zero. Taken
together, this results in the following system of equations for the forward
problem:
\begin{equation}
\left\{
\begin{array}{r@{\ }l@{\text{ at }}l}
\mathrm{div}\,\sigma(u) &= 0 & \Omega \setminus \mathcal F \text{ (static equilibrium, continuum)}\\
{}[\![ u ]\!] &= b \circ m^{-1} & \mathcal F \text{ (fault slip)} \\
{}[\![ \sigma(u) ]\!] \nu &= 0 & \mathcal F \text{ (static equilibrium across fault)} \\
\sigma(u) n &= 0 & \Gamma_{\rm surf} \text{ (traction-free surface)} \\
u &= 0 & \Gamma_{\rm far} \text{ (zero displacement at far field)} \\
\end{array}
\right.
\label{eqn:forward}
\end{equation}

Before looking into solution strategies for this system, it is readily apparent
that solutions to this problem are linear in $b$. Hence, for any manifold
$m \in \mathcal M$, there exists a linear map from the space of slip
distributions to the corresponding observations:
\begin{equation}
F_m : \mathcal B \rightarrow \mathcal D.
\label{eqn:forwardmap}
\end{equation}
If $\mathcal B$ and $\mathcal D$ are finite-dimensional and endowed with a
basis, we can identify every linear operator with a matrix, $F_m \in \mathbb
R^{\#\mathcal B \times \#\mathcal D}$, where $\#(\cdot)$ denotes the
cardinality of set $(\cdot)$.

\subsection{Analytical solutions}
\label{sec:fwd-analytical}

The general expression for solutions to \eqref{eqn:forward} was presented in
integral form by Volterra~\cite{Volterra_1907}:
\begin{equation}
u_n(y) = \sum_{ijkl} \int_{\xi \in \mathcal X} b_i(\xi) \nu_j(\xi)
C_{ijkl}(m(\xi)) \frac {\partial u_k^n} {\partial x_l}(m(\xi), y) \det\left|
\frac {\partial m} {\partial \xi} \right| d\xi,
\label{eqn:volterra}
\end{equation}
where the Green's function $u_k^n(x, y)$ is the $k$-th component of the
displacement vector in $x$ due to a unit point force at $y$ in direction $n$.
Note that \eqref{eqn:volterra} has an equivalent alternative form owing to the
symmetry relation $u_k^n(x,y) = u_n^k(y,x)$, which is a direct result of
Betti's reciprocal theorem~\cite{Love_1927}.

To obtain closed form expressions for the Green's function we need to place
additional constraints on our system.
Firstly, we require that the elastic properties of our medium are homogeneous
and isotropic, thus reducing the constitutive model to having two independent
parameters. Choosing as our parameters the Young's modulus $E$ and Poisson's
ratio $\nu$, the constitutive tensor becomes
\begin{equation}
C_{ijkl} = \frac E {2+2\nu} \left[ \frac {2\nu} {1-2\nu} \delta_{ij}
\delta_{kl} + \delta_{ik} \delta_{jl} + \delta_{il} \delta_{jk} \right].
\end{equation}
Since the displacement
resulting from a unit force is inversely proportional to the Young's modulus,
we observe that the displacement field resulting from Volterra's
equation~\eqref{eqn:volterra} depends only on the Poisson's ratio of the
medium.

Secondly, for closed form expressions for the Green's functions $u^k_n(x, p)$
to be available, we require the physical domain $\Omega$ to be a half space,
that is, to have an infinite, flat surface as its free surface $\Gamma_{\rm
surf}$ and the far field boundary $\Gamma_{\rm far}$ at infinity. The Green's
functions for the 2D half space were derived by Melan~\cite{Melan_1932}, those
for the 3D half space by Mindlin~\cite{Mindlin_1936}.

Building on Mindlin's results, closed form expressions for Volerra's equation
have been presented by Yoffe~\cite{Yoffe_1960} and Okada~\cite{Okada_1992},
subject to further restrictions in terms of fault planes and slip
distributions. While it is these results that are typically used in practical
applications, for the purposes of our study we shall apply Volterra's equation
directly so as not to incur additional and unnecessary restrictions to our case
studies that would diminish the value of the present study.

\subsection{The Weakly-enforced Slip Method}
\label{sec:fwd-wsm}

The Weakly-enforced Slip Method is a special case of the Finite Element Method,
which is in turn a Galerkin method, employing shape functions to construct a
finite system of equations that can be solved numerically. Contrary to
classical finite element treatments of Equation~\eqref{eqn:forward}, in which
the domain must be discretized such that the mesh conforms to the manifold $m$,
the defining property of the Weakly-enforced Slip Method is that the finite
element mesh can be formed independent of $m$.

Foregoing derivations and proofs, which are presented in detail in
\cite{vanZwieten_2014}, we present the WSM only in terms of its core result.
Given a finite element discretization for the computational domain $\Omega$
with mesh density $h$, and generating from it a discrete, vector-valued
function space $\hat V$, the WSM solution to Equation~\eqref{eqn:forward} is the
field $\hat u \in \hat V$ that satisfies, for all test functions $\hat v \in \hat V$,
\begin{equation}
\int_\Omega \nabla \hat u : C : \nabla \hat v =
\sum_{ijkl} \int_{\xi \in \mathcal X} b_i(\xi) \nu_j(\xi) \left\{
C_{ijkl}(m(\xi)) \frac {\partial \hat v_k} {\partial x_l}(m(\xi)) \right\}
\det\left| \frac {\partial m} {\partial \xi} \right| d\xi.
\label{eqn:wsm}
\end{equation}
Here $\{\cdot\}$ is the mean operator, which takes effect only in case $m$
coincides with an element boundary, making $\sigma(v_h)$ multi-valued; in the
general case it reduces to an evaluation. It is noteworthy that upon
substitution of the Green's function $u_k^n(m,y)$ for the test function $\hat
v_k(m)$, the left-hand-side reduces to $\hat u_n(y)$ and we obtain Volterra's
Equation~\eqref{eqn:volterra}.

The advantage of constructing the discrete solution space $\hat V$ independently
of $m$ is directly apparent from Equation~\eqref{eqn:wsm}: the stiffness
matrix, which results from the left hand side of the equation, as well as
solution primitives such as LU factors, are independent of $m$ and can thus be
created once and reused for many different faulting scenarios. The right hand
side vector, resulting from the right hand side of the equation, while
dependent of $m$, is constructed by integrating over the fault plane alone and
is therefore considerably cheaper to construct.

The disadvantage, as touched upon before, is that in constructing function
space $\hat V$ independently of $m$ it cannot possibly allow for
discontinuities at any subsequently defined manifold. By consequence,
displacement fields resulting from the WSM are continuous. Rather than a
distinct jump of magnitude $b$, the WSM solution exhibits a smeared out
transition local to the manifold. It was shown by \cite{vanZwieten_2014} that
the error thus incurred decreases exponentially with distance to the fault, and
that the method shows optimal convergence for any subdomain excluding the
manifold.

\section{Inverse Problem: Bayesian Formulation}
\label{sec:inverse}

The inverse problem is formulated in a Bayesian setting, the principles of
which can be found in, for instance, \cite[Ch.~1]{Tarantola_2005}. With
notation as introduced in Section~\ref{sec:forward}, we specify the main
problem statement as follows:

\textbf{(Inverse problem)}
\emph{Determine the probability distributions for the manifold $m \in \mathcal
M$ and slip distribution $b \in \mathcal B$ conditional to observations $d \in
\mathcal D$.}

Our stochastic framework consists of three random variables: manifold $M \in
\mathcal M$, slip distribution $B \in \mathcal B$, and line-of-sight surface
measurements $D \in \mathcal D$. The probability density of finding a manifold
$M = m$ and slip distribution $B = b$ given that we observe surface
displacements $D = d$ is given by Bayes' theorem as being proportional to the
likelihood of observing $D = d$ given $M = m$ and $B = b$, and the prior
probability of $M = m$ and $B = b$ absent observations:
\begin{equation}
\overbrace{f_{MB|D}(m,b,d)}^{\rm posterior} =
\overbrace{f_{D|MB}(d,m,b)}^{\rm likelihood}
\overbrace{f_{MB}(m,b)}^{\rm prior} /
\overbrace{f_D(d)}^{\rm marginal}.
\label{eqn:bayes}
\end{equation}

The marginal represents the probability of observing $D = d$. Its distribution
follows directly from the likelihood and the prior, owing to the fact that the
posterior probability density integrates to one. We present this relation for
completeness, though we will not need to evaluate it for our purposes:
\begin{equation}
f_D(d) = \int_{m \in \mathcal M} \int_{b \in \mathcal B} f_{D|MB}(d,m,b) f_{MB}(m,b)
\label{eqn:fD}
\end{equation}

This means that the only terms that require further elaboration are the prior,
the likelihood, and the posterior, which we shall explore in the following
sections.

\subsection{Prior Distribution}

The prior $f_{MB}(m,b)$ is the probability density of finding a manifold $m$
and fault slip $b$ absent any observations. It is a quantification of prior
knowledge of manifolds and slip distributions based on a general understanding
of physics as well as knowledge of the local tectonic setting.

A useful first step in the construction of our prior is to decompose it. Since
fault slip is defined on the manifold, a natural, universally valid
decomposition is the following:
\begin{equation}
f_{MB}(m,b) = f_M(m) f_{B|M}(b,m),
\end{equation}
where $f_M(m)$ is the prior probability density of the manifold,
and $f_{B|M}(b,m)$ the prior probability density of the fault slip conditional
to the manifold.

Before constructing a prior for the manifold we must identify $\mathcal M$ with
a parameter space. For instance, three coordinates, two angles and two
lengths define a rectangular plane. Additional parameters can encode
curvature, forks, or other irregularities as appropriate. Once defined, the
simplest prior is constructed by taking all parameters to be uncorrelated, and
every parameter either normally distributed around an expected value or
uniformly distributed within a chosen interval based on available in-situ
information. In case actual data is available about a-priori correlations and
distributions, this can directly be translated into a high quality prior.
If limited information is available, this can be encoded in a weakly
informative prior, e.g.{} a normal distribution with large standard deviation.

Depending on the regularity of $\mathcal M$ as defined by its parameterization,
it may be reasonable to assume that the prior probability distribution of the
slip does not depend much on the location, orientation, curvature, or other
properties of the fault. In other words, we may assume that the manifold and
fault slip are independent:
\begin{equation}
f_{B|M}(b,m) = f_B(b).
\end{equation}
Adopting this simplification, we further wish the slip vectors to be strongly
correlated at points that are close together, and weakly correlated at points
that are spaced far apart, based on a general understanding of the physics
underlying slip events. These notions are formalized in the positive
semi-definite autocovariance function $K: \mathcal X \otimes \mathcal X
\rightarrow \mathbb
R^{d-1 \times d-1}$:
\begin{equation}
K_{ij}(\xi_1, \xi_2) = \mathrm{cov}(B_i(\xi_1), B_j(\xi_2)),
\label{eqn:autocov}
\end{equation}
which we are free to design in any way that reflects existing prior knowledge.
Typically, $K(\xi_1,\xi_2)$ depends only on $\xi_1, \xi_2$ via their Euclidean
distance in relation to a specified correlation length.

For practical reasons we cannot operate on the infinite dimensional space
$\mathcal B$, but will instead operate on a finite dimensional subspace and
discrete random variables $\hat B \in \hat {\mathcal B} \subset \mathcal B$. To
aid its construction we define a vector-valued basis $h = \{h_1, h_2, \dots,
h_N\}$ for the discrete space of slip distributions, and thus associate with
any random slip $\hat B \in \hat {\mathcal B}$ a random vector $B^h \in \mathbb
R^N$ such that $\hat B(\xi) = \sum_{n=1\dots N} B^h_n h_n(\xi)$. We now take
$B^h$ to be normally distributed with covariance matrix $\Sigma_B^h$, which we
aim to construct in such a way that \eqref{eqn:autocov} still holds to good
approximation, i.e., $K(\xi_1, \xi_2) \approx h^T(\xi_1) \Sigma_B^h h(\xi_2)$.
To this end we multiply both sides of the equation by $h(\xi_1) h(\xi_2)^T$ and
integrate over the domain to form the following projection:
\begin{equation}
\underbrace{\int_{\xi_1 \in \mathcal X} \int_{\xi_2 \in \mathcal X} h(\xi_1)
K(\xi_1, \xi_2) h(\xi_2)^T}_{H_K} = \underbrace{\left(\int_{\xi_1 \in \mathcal
X} h(\xi_1) h(\xi_1)^T\right)}_{H_\delta} \Sigma_B^h
\underbrace{\left(\int_{\xi_2 \in \mathcal X} h(\xi_2)
h(\xi_2)^T\right)}_{H_\delta}
\label{eqn:kl}
\end{equation}
The extent to which the projection $\Sigma_B^h = H_\delta^{-1} H_K
H_\delta^{-1}$ approximates the autocovariance function depends on the details
of the autocovariance function $K$ in relation to the approximation properties
of the basis, but can in general be controlled fully by adding basis vectors,
i.e.\ increasing the dimension of $\hat {\mathcal B}$.

One remaining issue with the above construction is that the resulting
$\Sigma_B^h$ may not be positive semi-definite, which is a requirement for it to
qualify as a covariance matrix. We therefore proceed by diagonalizing the
result as $\Sigma_B^h = V \Lambda V^T$, where $\Lambda$ and $V$ are the
real-valued eigenvalues resp.\ $H_\delta$-orthogonal eigenvectors of the generalized
eigenvalue problem $H_K V = H_\delta V \Lambda$. Eliminating the negative
eigenvalues and corresponding eigenvectors we arrive at the covariance matrix
that approximates our autocovariance function $K$. In fact, we could go a step
further and eliminate small positive eigenvalues as well, as these modes are
seen to not contribute much to the overall expansion (more details on this
in Section~\ref{sec:slipbasis}) --- this process has the potential to greatly
reduce the dimension of $\hat {\mathcal B}$ and hence improve numerical
efficiency.

For the purposes of construction we were forced to make the difference explicit
between the true space of slip distributions, $\mathcal B$, and the finite
dimensional subspace that we will use for the analysis, $\hat {\mathcal B}$. We
note that, even though we did not need to make this formal, a similar
distinction applies to all spaces: our parametric space $\mathcal M$ is really
a finite dimensional subspace of the much larger space of possible manifolds,
and the observation space $\mathcal D$ is arguably a discrete subspace of a
continuous signal space. Since our analysis is finite dimensional, however, we
consider only (sufficiently rich) finite dimensional subspaces. For this reason
we shall also drop this distinction for slip distributions, and have $\mathcal
B$ denote the finite dimensional space going forward.

Finally, note that the covariance matrix is specific to the chosen basis $h$. We could
therefore still strive to create a basis $h'$ in such a way that the corresponding
covariance matrix $\Sigma_B^{h'}$ becomes an identity and all slip coefficients become
independent random variables. Indeed, the diagonalization provides us with the
tools we need in the form of the recombination matrix $V \Lambda^{1/2}$, post
removal of unwanted modes. The resulting basis is known as a Karhunen-Loeve
expansion \cite{Loeve_1977}, and it is what we shall be using in our practical
implementation. However, while we shall drop the suffix $h$ from here on, we
shall continue to write $\Sigma_B$, rather than $I$, in the interest of
preserving structure and keeping our derivations general.

\subsection{The Likelihood}

Given a manifold $m$ and slip distribution $b$, using a linear map $F_m$ of the
type of Equation~\eqref{eqn:forwardmap} we expect surface observations to equal
$F_m b$. Due to model errors and measuring noise, we take the likelihood of
observing $d$ to be normally distributed around this expected value with
covariance $\Sigma_D$:
\begin{equation}
f_{D|MB}(d,m,b) = G_{\Sigma_D}(d - F_m b),
\label{eqn:likelihood}
\end{equation}
with the Gaussian probability density function defined as
\begin{equation}
G_\Sigma(x) = \frac {\exp(-\tfrac12 x^T \Sigma x)} {\sqrt{\det|2\pi\Sigma|}}.
\label{eqn:gaussian}
\end{equation}
Since the sum of independent, normally distributed random variables is in turn
normal, the covariance matrix $\Sigma_D$ can be seen as the superposition of
several noise mechanisms. Spatially uncorrelated noise resulting directly from
the properties of the InSAR measurement system contributes to the diagonal,
with entries possibly varying to reflect dependence on distance or incidence
angle, or local factors such as caused by damage or other sources of temporal
decorrelation. Off-diagonal terms may be added to account for
spatially-correlated noise, such as errors caused by atmospheric delay.

Furthermore, though technically not noise, it is through the covariance that we
may account for the quality of the forward model itself. In the context of the
WSM we expect a large error in locations where the continuous solution space is
not able to follow local discontinuities. Making the variance locally large is
a convenient way of downweighing the data in this area, while the extreme case
of making it locally infinite effectively masks out the area, keeping only the
intermediate and far field data for the inversion.

\subsection{Posterior Distribution}

Substituting the prior probability distribution and the likelihood into Bayes'
theorem~\eqref{eqn:bayes}, we can rework terms to obtain the following result:
\begin{multline}
f_{MB|D}(m,b,d) = \overbrace{G_{\Sigma_D}(d - F_m b)}^{f_{D|MB}(d,m,b)} \overbrace{G_{\Sigma_B}(b)}^{f_B(b)} f_M(m) / f_D(d) \\[-3ex]
= \underbrace{G_{\Sigma_B'(m)}(b - b'(m,d))}_{f_{B|MD}(b,m,d)}
\overbrace{\underbrace{G_{\Sigma'_D(m)}(d)}_{f_{D|M}(d,m)} f_M(m) / f_D(d)}^{f_{M|D}(m,d)}
\label{eqn:posterior}
\end{multline}
with the posterior covariance and expected value of $B$ conditional to $M$ and $D$,
\begin{equation}
\Sigma'_B(m)^{-1} = \Sigma_B^{-1} + F_m^T \Sigma_D^{-1} F_m
\label{eqn:autocovslip}
\end{equation}
\begin{equation}
b'(m,d) = \Sigma'_B(m) F_m^T \Sigma_D^{-1} d.
\label{eqn:expectedslip}
\end{equation}
and the posterior covariance of $D$ conditional to $M$,
\begin{equation}
\Sigma'_D(m)^{-1} = \Sigma_D^{-1} - \Sigma_D^{-1} F_m \Sigma'_B(m) F_m^T \Sigma_D^{-1}
\end{equation}

The identity of Equation~\eqref{eqn:posterior} is verified through direct
substitution of the posterior covariances and expected value in the Gaussian
probability density function \eqref{eqn:gaussian}. Of particular use in this
exercise is the Weinstein-Aronszajn identity, $\det|I + A B| = \det|I + B A|$,
which, taking $A = F_m$ and $B = -\Sigma'_B(m) F_m^T \Sigma_D^{-1}$, results in
the following useful relationship:
\begin{equation}
\det|\Sigma_B| \det|\Sigma_D| = \det|\Sigma'_B(m)| \det|\Sigma_D'(m)|
\label{eqn:detident}
\end{equation}

While the identity of Equation~\eqref{eqn:posterior} is itself entirely
algebraic, the interpretation of the individual terms as conditional
probabilities $f_{M|D}$, $f_{B|MD}$ and $f_{D|M}$ is not immediately apparent.
The first follows from marginalizing over $\mathcal B$: since the marginal of
$G_{\Sigma_B'(m)}$ is 1 by definition, \eqref{eqn:posterior} directly leads to
the identity
\begin{equation}
f_{M|D}(m,d) = G_{\Sigma'_D(m)}(d) f_M(m) / f_D(d).
\label{eqn:fMD}
\end{equation}
Interpretation of the remaining conditional probabilities then follows readily
from the conditional probability relation $f_{B|MD}(b,m,d) = f_{MB|D}(m,b,d) /
f_{M|D}(m,d)$, and from Bayes' theorem, $f_{D|M}(d,m) = f_{M|D}(m,d) f_D(d) /
f_M(m)$.

The result of Equation~\eqref{eqn:fMD} is particularly useful as it allows us
to evaluate the total probability density of manifold $m$ conditional to
observations $d$, leaving the study of the slip $b$ to a separate, later stage.
While the expression contains the marginal $f_D$, which though a known quantity
is impractical to evaluate, this inconvenience is circumvented by using
sampling techniques that are insensitive to constant scaling. An example
of this is Markov Chain Monte Carlo (MCMC), which allows one to obtain low order
moments of the conditional probability density of $m$ using a feasibly low
number of evaluations.

Using any such techniques to single out a particular manifold, the probability
density of slip $b$ conditional to measurements $d$ and manifold $m$ is
normally distributed around expected value $b'(m,d)$ with posterior covariance
$\Sigma'_B(m)$. Interestingly, the latter is independent of the measurements,
meaning we can evaluate a-priori how a certain combination of measurement noise
properties, satellite viewing geometry and manifold position results in a
variance of the estimated slip along the length of the manifold.

In the expected value of Equation~\eqref{eqn:expectedslip} we recognize the
solution to a weighted least squares problem with Tikhonov regularization:
\begin{equation}
b'(m,d) = \argmin_{b \in \mathcal B} \left( \| F_m b - d \|_{\Sigma_D^{-1}}^2 +
\| b \|_{\Sigma_B^{-1}}^2 \right).
\end{equation}
While this is a standard method for solving ill-posed problems, the stochastic
interpretation thus obtained helps us in three ways. Firstly, it provides a
confidence measure of the result in the form of a posterior covariance matrix.
Secondly, it lends meaning to $\Sigma_B$ and $\Sigma_D$ that helps us to
construct the required matrices. And lastly, it enables one to design numerical
experiments that match the stochastic underpinnings of the method.

\section{Methodology}
\label{sec:method}

To test the WSM-based inversion of tectonic faulting, we synthesize a
deformation field for certain fault parameters $m$ and slip distribution $b$,
and then try to estimate the fault parameters $\hat m$ and slip distribution
$\hat b$ from noisy line-of-sight data using both the WSM and Volterra's
equation. The process can be divided in five steps:
\begin{enumerate}
\item Select fault parameters $m$ from $M$;
\item Draw a slip distribution $b$ from $B$ (or construct it manually);
\item Draw observation data $d$ from $D$ conditional to $M$ and $B$ using
Volterra's equation \eqref{eqn:volterra} as the forward model according to
Sections~\ref{sec:fwd-analytical} and \ref{sec:fwd-wsm};
\item Evaluate the posterior expected value $\hat m = E(M|D=d)$ and
covariances using either the WSM or Volterra's equation as the forward
model; \label{item:fwd}

\item[4\parbox{0pt}{*}.] Alternatively set $\hat m = m$ to study a linear-only inversion limited to $B$.
\item Evaluate the posterior expected value $\hat b = E(B|M=\hat m,D=d)$
and covariances using the same forward model as in \ref{item:fwd}.
\end{enumerate}
In the following we will elaborate on each of these steps.

\subsection{Constructing fault parameters}
\label{sec:faultparams}

Though any real world situation naturally concerns three-dimensional space,
it is advantageous to study a two-dimensional analogue as well as this allows
us to study the entire work flow in a setting that is less expensive and easier
to visualize. We will therefore construct two different sets of fault
parameters, one for one-dimensional faults in two-dimensional space, the other
for two-dimensional faults in three-dimensional space.

\begin{figure}
\centering
\begin{tikzpicture}[baseline=(current bounding box.center),scale=1.2]
\draw (-2,0)--(2,0)--(2,-2)--(-2,-2)--cycle;
\draw [thin,<->] (-2,-2.1)--(2,-2.1); \draw (0,-2.1) node [below] {\scriptsize 100\,km};
\draw [thin,<->] (2.1,-2)--(2.1,0); \draw (2.05,-1) node [right] {\scriptsize 50\,km};
\draw [thin,dashed] (-.4,0)--(-.4,-.5);
\draw [thin,<->] (-.5,-.02)--(-.5,-.47); \draw (-.45,-.25) node [left] {\scriptsize depth};
\draw (0,0) node [above right] {\scriptsize free surface};
\draw [fill=white] (0,0) circle (.05);
\draw [|->,>=stealth] (0,-.17)--++(-.38,0); \draw (-.36,-.33) node [right] {\scriptsize position};
\begin{scope}[shift={(-.4,-.5)}] 
  \fill (0,0) circle (.05);
  \draw [ultra thick] (0,0) -- +(-120:1.5);
  \draw [thin,dashed] (-1,0) -- (2,0);
  \draw [thin] (-.5,-.05) arc (0:75:-.3);
  \draw (-.4,-.1) node [below left] {\scriptsize dip angle};
  \draw [thin,<->] (-30:.1)--++(-120:1.5); \draw (-30:.1)--++(-120:.75) node [right] {\scriptsize size = 50km - depth};
\end{scope}
\begin{scope}[shift={(1.2,-.5)}] 
  \fill [lightgray] (0,0) circle (.05);
  \draw [ultra thick,lightgray] (0,0) -- +(-45:1.5);
  \draw [thin] (0,0) ++(-45:.5) -- ++(80:.25);
  \draw [thin] (0,0) ++(-45:.5) -- ++(80:-.25);
  \draw [thin] (0,0) ++(-45:.5) -- ++(190:.25);
  \draw [thin] (0,0) ++(-45:.5) -- ++(190:-.25);
\end{scope}
\end{tikzpicture}
\hspace{-2cm}\hfill
$\left. \begin{array}{r} \text{2D} \left\{\begin{array}{r@{}}
\text{position} \\
\text{dip angle} \end{array} \right. \\ \begin{array}{r@{}}
\text{offset} \\
\text{strike angle} \end{array} \end{array} \right\} \text{3D}$
\begin{tikzpicture}[baseline=(current bounding box.center),scale=1.2]
\draw (-2,-.5)--(1,-.5)--(1,-2)--(-2,-2)--cycle;
\draw (-2,-.5)--(-1,0)--(2,0)--(2,-1.5)--(1,-2);
\draw (1,-.5)--(2,0);
\draw [dotted] (-2,-2)--(-1,-1.5)--(-1,0);
\draw [dotted] (-1,-1.5)--(2,-1.5);
\draw [thin,<->] (-2,-2.1)--(1,-2.1); \draw (-.5,-2.1) node [below] {\scriptsize 100\,km};
\draw [thin,<->] (1.05,-2.1)--(2.05,-1.6); \draw (1.5,-1.75) node [below right] {\scriptsize 100\,km};
\draw [thin,<->] (2.1,-1.5)--(2.1,0); \draw (2.05,-.75) node [right] {\scriptsize 50\,km};
\draw (0,0) node [above] {\scriptsize free surface};
\draw [->,>=stealth] (0,-.25)--(.4,-.15)--(.3,-.35);
\draw (.34,-.14) node [right] {\scriptsize position};
\draw (.3,-.33) node [right] {\scriptsize offset};
\draw [fill=white] (0,-.25) circle (.05);
\draw [thin,dashed] (.3,-.85)--++(0,.5);
\draw [thin,<->] (.2,-.83)--++(0,.48); \draw (.28,-.63) node [left] {\scriptsize depth};
\begin{scope}[shift={(.3,-.85)}] 
  \fill (0,0) circle (.05);
  \draw [ultra thick] (-.1,-.2) -- ++(.2,.4) -- ++(.5,-.8) -- ++(-.2,-.4) -- cycle;
  \draw [thin,<->] (.1,.2)++(.05,.1) -- ++(.5,-.8); \draw (.4,-.14) node [right] {\scriptsize size};
  \draw [thin,<->] (-.1,-.2)++(.5,-.8)++(.0625,-.1) -- ++(.2,.4); \draw (.5,-.88) node [right] {\scriptsize size};
  \draw [thin,dashed] (0,0) -- ++(.5,-.8);
  \draw [thin,dashed] (-.8,0) -- (.8,0);
  \draw [thin,dashed] (0,0) -- ++(-.8,-.2);
  \draw [thin,dashed] (0,0) -- ++(.8,.2);
  \draw (.02,.01)++(.1,.06)++(.2,.05)--++(.05,.1)--++(-.2,-.05);
  \draw [thin] (-.5,0)..controls (-.7,-.1) and (-.45,-.2)..(-.05,-.1); \draw (-.3,-.3) node [left] {\scriptsize strike angle};
  \draw [thin] (-.3,-.05) arc (0:130:-.3); \draw (.08,-.5) node [left] {\scriptsize dip angle};
\end{scope}
\end{tikzpicture}
\caption{Fault parameters and their representation in 2D and 3D scenarios. The
invalid parameter combination shown in gray in the 2D scenario is not a member
of $\mathcal M$.}
\label{fig:faultparams}
\end{figure}
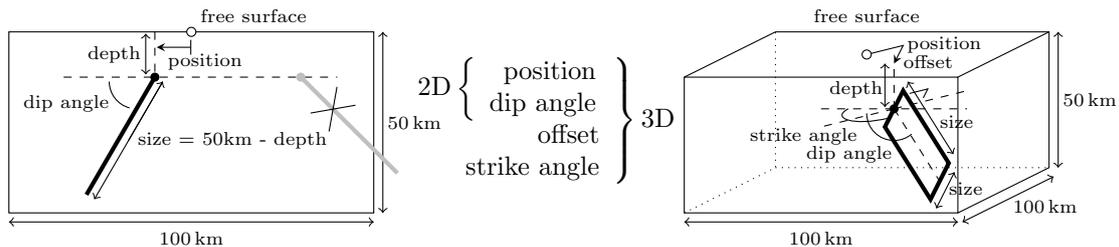

We limit ourselves to the space of straight faults of fixed dimensions, that
are placed anywhere in a box of 100 km width, 100 km breadth (for 3D scenarios)
and 50 km depth. In order to distinguish between the class of rupturing and
non-rupturing faults (a distinction that can often be made on the basis of
field observations) we set the minimum depth to remain fixed. This leaves two
parameters in 2D space and four parameters in 3D space to parameterize the
entire space, as summarized in Figure~\ref{fig:faultparams}.

We note that while the fixed dimensions of the fault plane form an upper bound
for the dimensions of the fracture zone, the support of the slip distribution
can still localize within these confines. Introducing additional parameters for
length and width would therefore not add actual degrees of freedom but rather
ambiguities between the two spaces $\mathcal M$ and $\mathcal B$, manifested in
additional expenses for the nonlinear inversion due to the increased dimension
of $\mathcal M$. While omitting dimensions from the parameter space means that
we require expert judgement to define what size is sufficiently large for a
given situation, we can verify the validity of this assumption a-postiori, for
example by testing if the inverted slip is sensitive to fault plane
enlargement.

Similar considerations apply to fault location, where in-plane variations can
to some degree be captured by $\mathcal B$. This is what allows us to fix the
depth, while the actual onset of slip might be deeper still. Relatedly, in the
3D scenario we encode the location as a `position' that is normal to strike,
and an `offset' along strike. While we could conceivably eliminate the
latter and rely entirely on $\mathcal B$ to capture the in-plane component, we
choose to keep the offset in $\mathcal M$, as the fault plane would otherwise
have to be undesirably large in order to still cover the search box in all
orientations. However, we anticipate that its posterior variance will be
significantly larger than that of the position due to the remaining ambiguities
between $\mathcal M$ and $\mathcal B$.

The fault size is the largest size that fits the box given the vertical offset.
Rupturing faults are thus of 50 km length, while faults that close 10 km below
the surface are of 40 km length. For the 3D scenario both dimensions of the
fault are always kept equal. The mapping to physical space $m(\xi)$ is an
affine transformation, supporting the assumption that the slip distribution
$b(\xi)$ can be considered independently. As it is for our purposes important
that the entire fault fits inside the search box, we define $\mathcal M$ to
contain no positions that fall outside of it, nor any angle that causes the
fault to intersect its boundary. Figure~\ref{fig:faultparams} shows an example
of this in the 2D scenario. On the resulting oblique domain we take the prior
distribution of $M$ to be uniform.

\subsection{Constructing the slip distribution}
\label{sec:slipbasis}

To construct a space of slip distributions we take the local fault plane
coordinates to be the unit line or unit square, $\mathcal X = [0,1]^{d-1}$,
mapping through $m$ onto a fault plane of dimensions $L$. We take slip
components in orthogonal directions to be independent, which is a
non-restrictive assumption in practice. This reduces the distribution of the
vector field to a series of identically distributed scalar fields, for which we
define an exponential autocorrelation function with correlation length $\ell$
and slip amplitude $\beta$:
\begin{equation}
K(\xi_1, \xi_2) = \beta^2 w(\xi_1) w(\xi_2) \exp\left(-\tfrac12 \tfrac {|\xi_1
- \xi_2|^2} {(\ell/L)^2} \right).
\end{equation}
Here $w$ is a window function. The window sets the variance to zero on all
boundaries for non-rupturing faults, or for all but the surface edge for
rupturing faults. In 2D space (with a 1D fault) these window functions are
$w_{\rm closed}(\xi) = 4 \xi (1-\xi)$ and $w_{\rm open}(\xi) = \xi (2-\xi)$,
respectively. In 3D space (with a 2D fault) they are the tensor product of
$w_{\rm closed} \otimes w_{\rm closed}$ and $w_{\rm closed} \otimes w_{\rm
open}$.

A Karhunen-Loeve expansion is constructed for this autocovariance function via
the projection of Equation~\eqref{eqn:kl} using a truncated trigonometric
series for the basis $h$ --- though we remark that a mesh-based construction
can be used instead in case more flexibility is required. Similar to the window
functions, we distinguish the non-rupturing and the rupturing situations. For
non-rupturing faults in one dimension, we use the orthonormal sine series $
h_n(\xi) = \sqrt{2} \sin(\xi n \pi) $. For rupturing faults we use a modified
cosine series $ h_n(\xi) = \sum_{i=0\dots n} \alpha_{ni} \cos(\xi i \pi) $,
with coefficients $\alpha_{ni}$ chosen such that $h_n(1) = 0$ and the basis
functions are orthonormal. For two-dimensional faults we use the outer products
to form a scalar basis on the unit square, preserving orthonormality.

\begin{figure}
\centering
\includegraphics[scale=.67,trim=5mm 0mm 10mm 10mm]{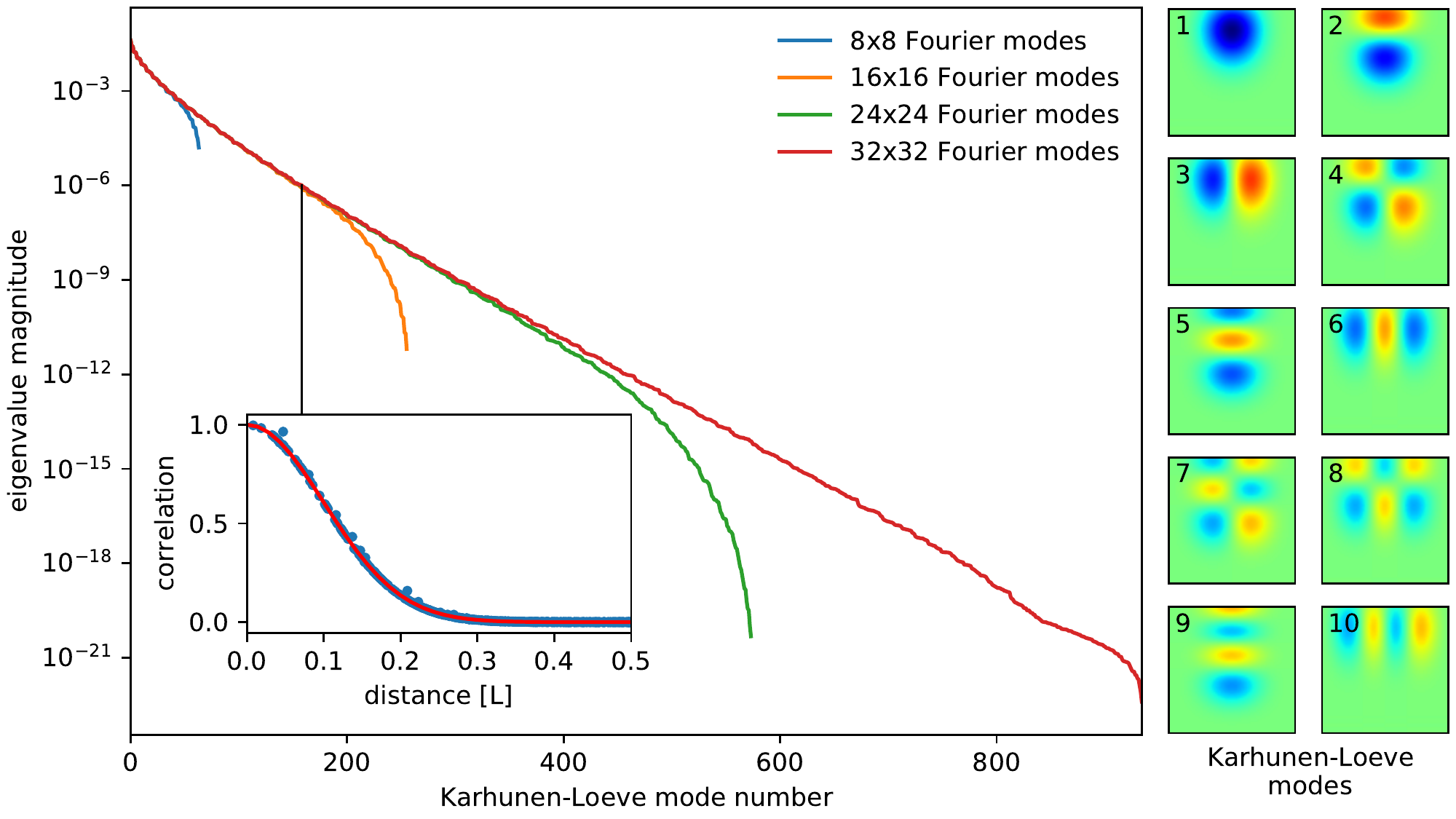}
\caption{The eigenvalue spectrum $\Lambda$ of the $H_K$ matrix for a rupturing
fault in three-dimensional space and a correlation length $\ell = 0.1 L$,
computed at different truncation points of the trigonometric series. The inset
shows the autocorrelation of the 153-mode Karhunen-Loeve expansion based on
1024 randomly selected point pairs (blue) along with the target autocorrelation
function (red). Shown to the right are the first 10 Karhunen-Loeve modes $V
\Lambda^{1/2}$.}
\label{fig:klmodes}
\end{figure}

While orthonormality is not a requirement, it is a convenient property as we no
longer need to form $H_\delta$ (now an identity) and the generalized eigenvalue
problem reduces to a conventional eigenvalue problem $H_K V = V \Lambda$ ---
the size of which depends on the truncation point of the trigonometric series.
Selecting the $n$ largest eigenvalues, the Karhunen-Loeve expansion is formed
by recombining the modes by $V \Lambda^{1/2}$ causing the corresponding
covariance matrix to reduce to an identity; drawing a sample from the
distribution then amounts to independently drawing $n$ coefficients from a
standard normal distribution. Since orthogonal slip components are taken to be
independent in our choice of autocovariance function, it suffices to form a
scalar basis for each of the components of the vector. Figure~\ref{fig:klmodes}
shows the spectrum $\Lambda$ for first nine scalar bases functions for
illustration.

For our experiments we set the correlation length to $\ell = 5$ km and the slip
amplitude to $\beta = 1$ m. These values are representative of actual
geophysical conditions but otherwise arbitrary. To determine the number of
Karhunen-Loeve modes to retain, we set a maximum error of $\| H_K - V \Lambda
V^T \| <$ 10 mm$^2$ measured in the Frobenius norm. By this process we
establish that $n=12$ Karhunen-Loeve modes are sufficient for the 2D scenario
and $n=153$ modes for the 3D scenario. Repeating this process for different
truncation sizes of the original trigonometric series we find that this result
is stable beyond 16 modes per dimension, and decide based on this that 32 basis
functions per dimension is a sufficiently rich starting point for the
expansion.

In a practical application one may wonder if the constructed prior distribution
is an accurate representation of reality, and indeed if the true slip
distribution is even an element of $\mathcal B$ at all. To make sure that a
slip distribution drawn from the prior does not represent an artificial best
case scenario with little real world value, we additionally construct a slip
distribution that has local support at a select area of the fault. In addition
to testing the robustness of the method to incorrect assumptions, the local
support also allows us to test whether fault dimensions can indeed be captured
via the slip, rather than via additional fault parameters.

\subsection{Synthesizing observation data}
\label{sec:synth}

We synthesize a displacement field for a given fault $m$ and slip distribution
$b$ based on the assumption that that the free surface is flat and infinite,
and the material properties are homogeneous. In this situation we have a
fundamental solution available in the form of Melan's (2D) and Mindlin's (3D)
solution, which means we can synthesize the displacement field by evaluating
Volterra's Equation~\eqref{eqn:volterra}. The integral is evaluated numerically
by means of Clenshaw-Curtis quadrature up to a truncation error that is well
below the selected noise level.

\begin{figure}
\centering
\includegraphics[scale=.67,trim=8mm 12mm 12mm 19mm,clip]{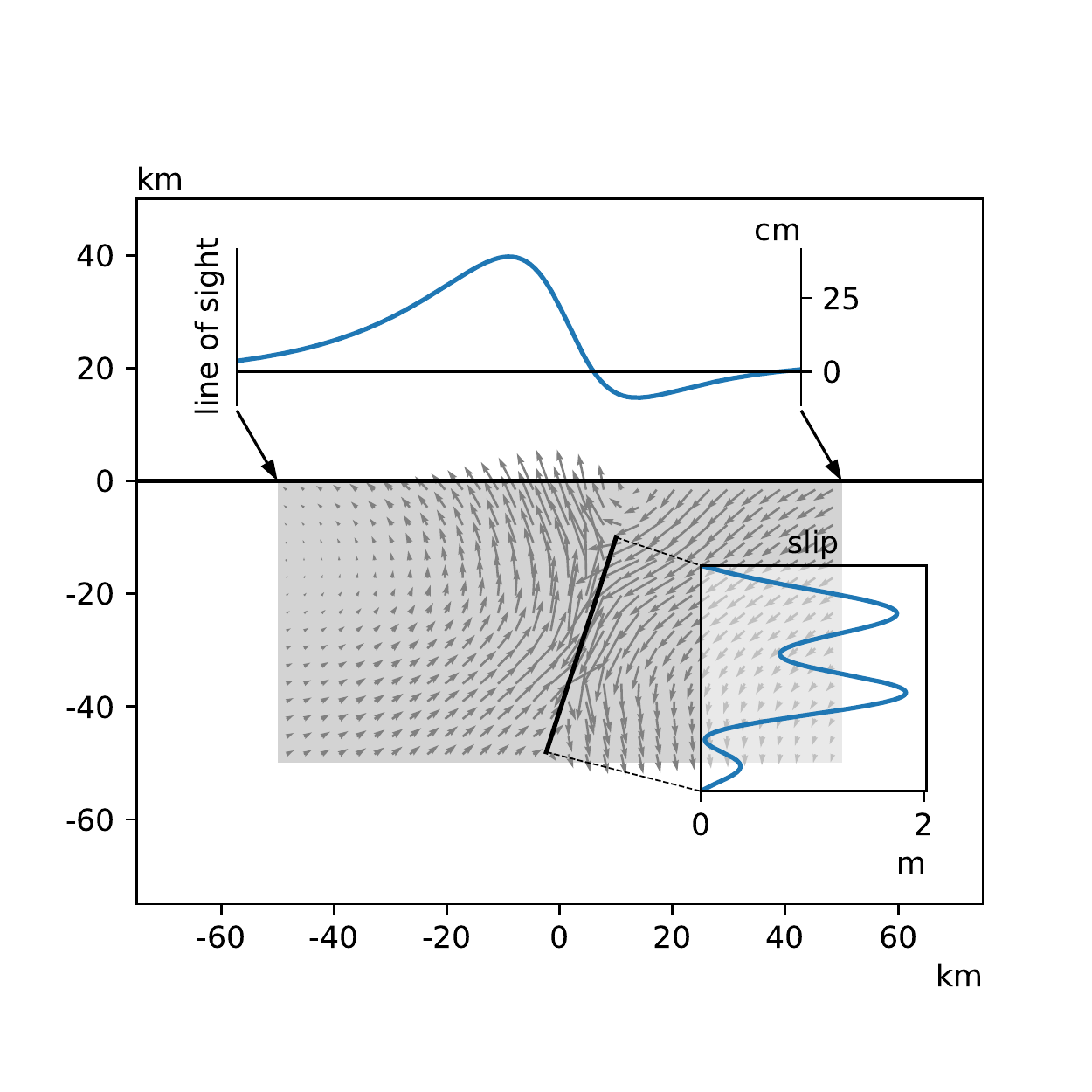}\hfill%
\includegraphics[scale=.67,trim=8mm 12mm 12mm 19mm,clip]{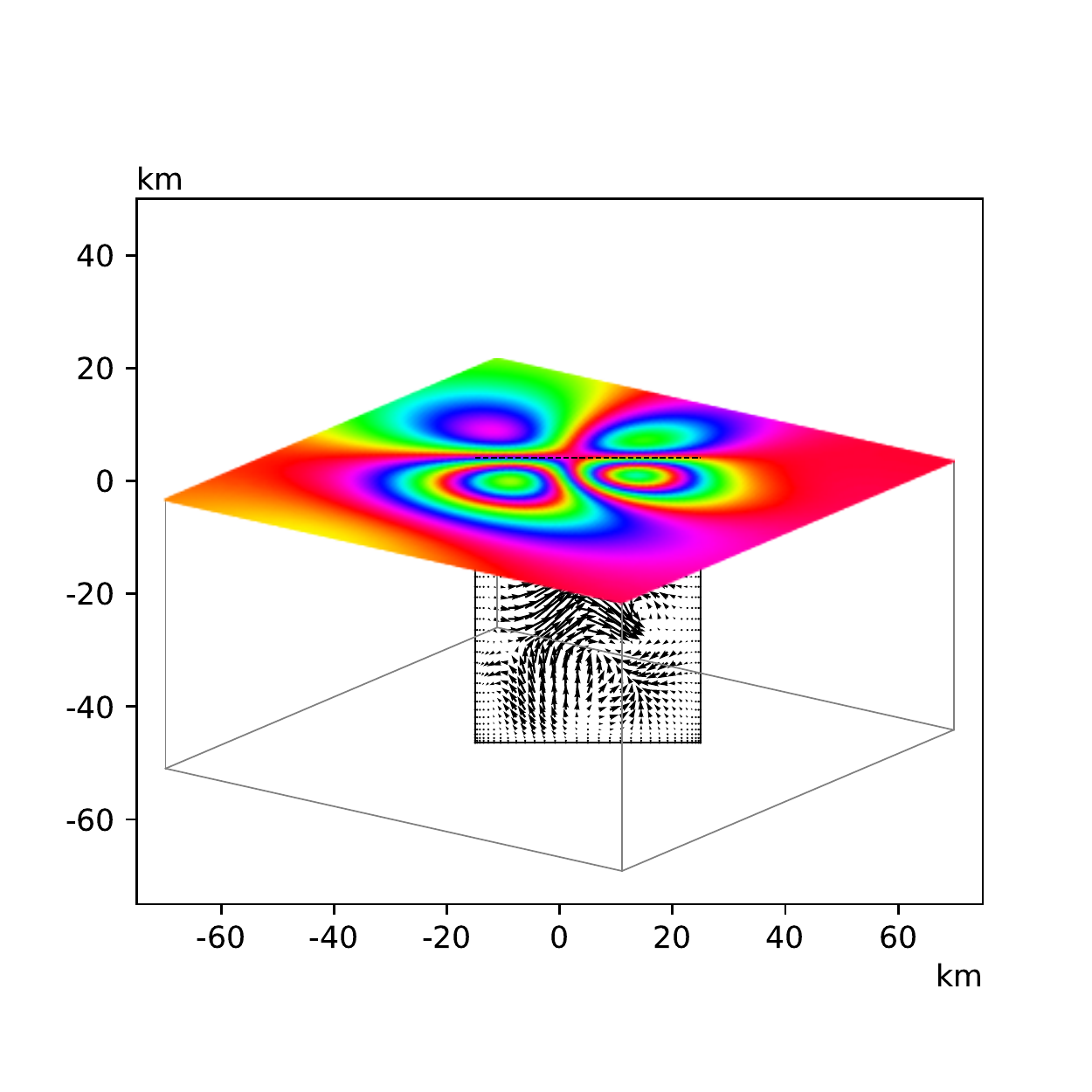}
\caption{Visualisation of the search box containing a non-rupturing fault at
position 10km ($0.2L$) and dip angle $72^\circ$ ($0.4\pi$), and in 3D
additionally offset 5km ($0.1L$) and strike angle $54^\circ$ ($0.3\pi$)
(Section~\ref{sec:faultparams}), a slip distribution drawn from the prior
distribution (Section~\ref{sec:slipbasis}) and corresponding synthesized
deformation data at a $30^\circ$ incidence angle (Section~\ref{sec:synth}). Left: a
2D scenario. Right: a 3D scenario in orthographic projection normal to the
fault, showing line of sight deformations modulo 2.8 cm to mimic a typical
C-band interferogram.}
\label{fig:situation}
\end{figure}

We fix Poisson's ratio at $\nu=0.25$ for all experiments. The displacement
field is evaluated in a uniformly spaced grid at a 50 meter resolution,
covering the top of the search box defined in Section~\ref{sec:faultparams}.
The displacements are then projected onto a vector at a 30 degrees incidence
angle, simulating line-of-sight observations as obtained from a typical
satellite mission. Finally, Gaussian noise is added to the line-of-sight data
that is spatially uncorrelated (thus ignoring atmospheric delays) and normally
distributed with a standard deviation of 1 mm. A typical result can be seen in
Figure~\ref{fig:situation}. Note that, while the displacement is displayed
without noise, the assumed noise is so small relative to the magnitude of the
signal that adding it would not make a visible difference.

A difference between the raw synthesised data and deformation data received
from a satellite mission is that the SAR sensor provides deformation data
modulo the sensor's semi-wave length (for details see
e.g.~\cite{Hanssen_2001}), measuring $d + k \lambda/2$ for an unknown integer
value of $k$. While this is typically solved through phase unwrapping (under
the assumption that $|\Delta d| < \lambda/4$ for adjacent points) it leaves the
data inherently lacking an absolute reference. As we aim for this synthesized
study to be representative of real world situations, we need to make sure that
our measurement data is similarly relative. To this end we select one
measurement point as a reference and subtract its deformation from that of the
other measurement points. The inversion is then performed based on the
differenced data, using a forward model that reflects the identical
differencing procedure.

To formalize this procedure we introducing the unit vector $r \in \mathbb R^n$
that selects the reference point, and $e \in \mathbb R^n$ the vector of ones.
With that we can express the differencing operator as
\begin{equation}
D = I - e r^T.
\end{equation}
The differencing is followed by a restriction operation that removes the
reference point from the data, for which we introduce the operator $R \in
\mathbb R^{n \times n-1}$. Note that $r$ and $R$ are related via $r^T R = 0$
and $R R^T + r r^T = I$.

Regardless of the covariance of the measurement noise, all differenced data is
correlated due to the shared reference point: $ E((d_i - d_{\rm ref}) (d_j -
d_{\rm ref})) \neq 0$ in general due to the nonzero variance of $d_{\rm ref}$.
Using the differencing and restriction operators we can express the covariance
matrix of the differenced data in terms of that of the original measurements,
which we shall denote henceforth as $\check\Sigma_D$:
\begin{equation}
\Sigma_D = R^T D \check\Sigma_D D^T R.
\end{equation}
The additional covariances render the matrix fully dense. Fortunately, we note
that we can express the inverse covariance of the differenced data in terms of
that of the original measurements. Therefore, if the original covariance matrix
can be inverted efficiently (for instance if it is diagonal) then this property
carries over to the new covariance matrix:
\begin{equation}
\Sigma_D^{-1} = R^T X \check\Sigma_D^{-1} R,
\label{eqn:invsigmaD}
\end{equation}
where $X$ is a rank-1 update defined as
\begin{equation}
X = I - \frac {\check\Sigma_D^{-1} e e^T} {e^T \check\Sigma_D^{-1} e}.
\label{eqn:invdiffcov}
\end{equation}
While our choice of spatially uncorrelated noise implies that the covariance
matrix is diagonal, \eqref{eqn:invsigmaD} holds for any $\check\Sigma_D$ and is easily
verified using the identities $D^T X = X$, $D X^T = D$ and $X
\check\Sigma_D^{-1} = \check\Sigma_D^{-1} X^T$, together with $D R = R$, $R R^T
D = D$ and $R^T R = I$. Using the same identities it further follows that
\begin{equation}
D^T R \Sigma_D^{-1} R^T D = X \check\Sigma_D^{-1}.
\label{eqn:XSigmaD}
\end{equation}

This last result \eqref{eqn:XSigmaD} is noteworthy for two reasons. Firstly, in
Equations~\eqref{eqn:autocovslip} and \eqref{eqn:expectedslip} the inverse
covariance occurs only surrounded by either the forward model or the data, both
of which need to be differenced as $F_m = R^T D \check F_m$ and $d = R^T D
\check d$. The result shows that it is not necessary to perform these
operations explicitly, and that a rank-1 update of the original covariance
matrix inverse is all it takes to switch from an inversion of absolute
measurements to that of relative measurements. Secondly, the absence or $R$ and
$r$ in \eqref{eqn:invdiffcov} proves that the inversion is entirely insensitive
to the chosen reference point --- indeed, we do not need to make any choice at
all.

\subsection{Sampling the posterior distribution}

We are interested in evaluating the expected value and auto-covariance of
$M|D$, i.e.{} the posterior distribution of fault parameters given the measurements at
the surface. The probability density function $f_{M|D}(m,d)$ was presented in
\eqref{eqn:fMD}. However, evaluation of this expression for a
particular value of $m$ is problematic because of the marginal $f_D(d)$
contained within, which, while defined in Equation~\eqref{eqn:fD}, does not
have a closed form expression. As such we cannot feasibly evaluate the Lebesgue
integral to compute the desired quantities.

We find a solution in the class of Markov Chain Monte Carlo (MCMC) methods,
which provides an algorithm for drawing samples from $M|D$ while relying only
on the ratio of the probability density function at two points $m_1$ and $m_2$,
thereby cancelling the marginal and other factors that are independent of $m$.
As the sample sequence $\{m_1, m_2, \dots\}$ thus produced has an empirical
probability measure that coincides with the posterior distribution, the
expected value and higher moments can be evaluated using Monte Carlo
integration
\begin{equation}
E(g(M|D)) = \lim_{N\rightarrow \infty} \frac1N \sum_{i=1\dots N} g(m_i),
\label{eqn:mcintegral}
\end{equation}
for any $g$, which we can truncate at any $N$ depending on the desired level of accuracy.

While Equation~\eqref{eqn:fMD} shows $f_{M|D}(m,d)$ to depend on $f_M(m)$, we
stated in Section~\ref{sec:faultparams} that we take the prior distribution of
$M$ to be uniform. As such the probability density function is proportional
only to $\mathcal N_{\Sigma_D'(m)}(d)$. Writing out the multivariate normal
distribution and reworking terms we obtain the following identity:
\begin{equation}
\mathcal N_{\Sigma_D'(m)}(d) = 
\frac {\exp(-\tfrac12 d^T \Sigma_D^{-1} d)} {\sqrt{(2\pi)^{\#\mathcal D} \det|\Sigma_B \Sigma_D|}}
f_d(m), \quad
f_d(m) = \frac {\exp(\tfrac12 d^T \Sigma_D^{-1} F_m b'(m,d))} {\sqrt{\det|\Sigma_B'(m)^{-1}|}}.
\label{eqn:mcmcfunc}
\end{equation}
Here we made use of identity \eqref{eqn:detident} to rewrite
the determinant of the potentially very expensive covariance matrix
$\Sigma_D'(m)$ as the determinant of the smaller and less complex matrix
$\Sigma_B'(m)^{-1}$. We also isolated a proportionality constant independent of
$m$, leaving only $f_d(m)$ to feature in our MCMC method.

Every evaluation of the resulting function involves a linear inversion of the
slip for given $m$, as is seen directly from the presence of the posterior
expected value $b'(m,d)$ and covariance matrix $\Sigma_B^{-1}(m)$. The
construction of these and of the forward model $F_m$ will be discussed in
Section~\ref{sec:slipinv}, which details the linear slip inversion process.
Of note presently is that when the required inversion of $\Sigma_B^{-1}(m)$ is
performed via a Cholesky decomposition, then the trace of the Cholesky matrix
conveniently equals the square-root determinant in
Equation~\eqref{eqn:mcmcfunc}.

The particular MCMC method selected for our purpose is the
Metropolis-Hastings~\cite{Hastings_1970} algorithm, which performs a random
walk through the sample space $\mathcal M$ using a proposal distribution to
generate candidates, combined with an acceptance/rejection step based on the
ratio of probability densities. The algorithm as it is employed here consists
of the following steps:
\begin{enumerate}
\item initialize $m_0 \in \mathcal M$
\item for $i = 1, 2, \dots, N$:
\begin{enumerate}
\item draw a random update vector $\delta_m \in \mathcal M$ from the proposal distribution
\item draw a uniform random number $u \in [0,1]$
\item set $m_{i} = m_{i-1} + \begin{cases} \delta_m & \text{ if $f_d(m_{i-1} + \delta_m) \geq u f_d(m_{i-1})$ (accept update)} \\ 0 & \text{ otherwise (reject update)} \end{cases}$ \label{mh:update}
\end{enumerate}
\end{enumerate}
Note that condition of step~\ref{mh:update} implies that the update is always
accepted if it leads to a state of higher probability. Note further that if an
update is rejected the current state is repeated in the sequence, thereby
adding weight to the empirical distribution and subsequent Monte Carlo
integration \eqref{eqn:mcintegral}.

What remains is to define the starting vector and the proposal distribution.
While the proven convergence of the Markov chain suggests that both can be
chosen arbitrarily if we take $N$ sufficiently large, this approach is not
feasible in practice as the number of iterations required to escape from a
local maximum can be prohibitively large. Instead, we require the starting
point to be reasonably close to where $f_d$ takes its maximum, and the proposal
distribution to be locally similar to $f_d$ in order to have a reasonable
acceptance rate.

Aiming for the global maximum, we select the starting vector $m_0$ using a grid
search to find local maxima followed by the Nelder-Mead uphill simplex method.
The proposal distribution is taken to be Gaussian with covariance $\Sigma_P$,
which we would like to resemble the distribution of $f_d$ local to $m_0$.
Aiming to use a projection in logarithmic space, we wish to form a symmetric
matrix $A$ such that, for all $\hat m$ in the vicinity of $m_0$,
\begin{equation}
-\tfrac12 (\hat m - m_0)^T A (\hat m - m_0) \approx \log f_d(\hat m) - \log f_d(m_0).
\end{equation}
As this relation is linear in the matrix coefficients $A_{ij}$ we can optimize
it using the weighted linear least squares method, in which we reuse the
sequence $\{ \hat m_i \}$ of Nelder-Mead iterates as data points, and
$f_d(\hat m_i)$ as weights in order to downweigh the tails of the
distribution. We note that, while it is convenient to reuse available
data, we are at liberty to augment the series with extra evaluations in the
vicinity of the optimum to increase the quality of the projection, even though
we have found no need to do so for the cases considered.

Finally, we use the optimal scaling result of Roberts et al~\cite{Roberts_1997} to 
form the covariance matrix of our proposal distribution,
\begin{equation}
\Sigma_P = \frac {(2.38)^2} {\#\mathcal M} A^{-1}.
\end{equation}

\subsection{Evaluating the posterior expected slip}
\label{sec:slipinv}

The expected value $E(B|MD)$ of the slip distribution given a fault $m$ and
surface measurements $d$ is provided in closed form by
Equation~\eqref{eqn:expectedslip}. The posterior covariance matrix is defined
in Equation~\eqref{eqn:autocovslip}, in which $\Sigma_B$ is the identity matrix
owing to the properties of the Karhunen-Loeve expansion. Crucially, both
involve the formation of the forward model $F_m$, which maps the coefficient
vector that encodes the slip distribution onto the corresponding vector of
surface deformation gradients. It follows that the rows of the matrix $F_m$ are
formed by the surface deformations corresponding to the slip distribution that
is represented by the individual basis vectors $h_n(\xi)$ that we constructed
in Section~\ref{sec:slipbasis}.

If the selected forward model is Volterra's equation then $F_m$ is formed by
repeated evaluation of Equation~\eqref{eqn:volterra}. If the selected model is
the WSM then constructing $F_m$ involves constructing a finite element matrix
and solving it for a block of right-hand-side vectors. Having discussed the
evaluation of Volterra's equation in Section~\ref{sec:synth}, we will use the
remainder of this section to elaborate on details of the latter.

A first step in any finite element computation is the formation of the
computational mesh on which the discrete basis is formed, in our case to
describe the displacement field. Recall from Section~\ref{sec:faultparams} that
all fault planes $\mathcal M$ will be confined in a rectangular box of given
size. We now create a regularly spaced grid of elements spanning this search
box, allowing us to cheaply trace any physical coordinate inside the box to the
containing element and its element-local coordinate which will greatly aid the
efficiency of the fault plane integration. We note, however, that efficient
lookup procedures exist for other mesh types as well, for example using quad
trees~\cite{Krause_1996} or alternating digital trees~\cite{Bonet_1991}.

Since our computational domain is a halfspace we have no boundary conditions to
place on the walls of the search box, except for the free surface which is
traction free. Instead we take the infinite element approach of extending our
mesh with several rows of extra elements and using a geometric map to
continuously stretch the elements outside the search box towards infinity.
Specifically, if $2L$ is the width of the box, $2n_{\rm box}$ the number of
elements spanning the box and $2n_{\rm inf}$ the number of elements spanning
infinity, we apply the following piecewise hyperbolic map to every spatial
dimension:
\begin{equation}
x_i(e) = \frac L {n_{\rm box}} \left\{\begin{array}{ll@{\ }l}
e - (n_{\rm box} + e)^2 / (n_{\rm inf} + e) & -n_{\rm inf} &< e < -n_{\rm box} \\
e & -n_{\rm box} &\leq e \leq +n_{\rm box} \\
e + (n_{\rm box} - e)^2 / (n_{\rm inf} - e) & +n_{\rm box} &< e < +n_{\rm inf}
\end{array}\right.
\label{eqn:mesh}
\end{equation}
Note that this includes the depth direction, in which case we take $-n_{\rm inf} < e
\leq 0$. Unless stated otherwise we will select a infinity-to-box ratio of
$n_{\rm inf}/n_{\rm box} = \tfrac32$, which means that in 3D the treatment of the far field increases
the number of elements by a factor $(\tfrac32)^3 \approx 3.38$ relative to the
number of elements in the search box.

In creating the discrete function space $V_n$ we make use of the fact that our
mesh is structured by creating a C$^1$ quadratic spline basis, also known as
isogeometric analysis~\cite{Hughes_2010}, which we showed in
\cite{vanZwieten_2014} to have better accuracy to degrees of freedom, and we
remove the outermost basis functions to impose the far field constraint. With
that we are in a position to evaluate the left hand side of
Equation~\eqref{eqn:wsm} to form the stiffness matrix. Since this matrix will
be reused many times we also invest the time to construct a high quality
preconditioner, opting, in fact, to form a complete Cholesky decomposition.

The right-hand side of Equation~\eqref{eqn:wsm} involves an integral over
$\mathcal X$ for which we use the same Clenshaw-Curtis quadrature scheme that
we used for the synthetization of observation data in Section~\ref{sec:synth},
while making use of the rectilinearity of the mesh in the search box to locate
the corresponding element coordinates necessary to evaluate the basis
functions. Having formed the right-hand side vector we can solve the system and
form the discrete solution $u_h \in V_n$, which can then be evaluated in any
point of our choosing.

\section{Results}
\label{sec:results}

We will present several results of the process outlined in
Section~\ref{sec:method}, with the aim of illustrating the many variables and
their effect on the overall computation. While we are mainly interested in
applications in three-dimensional space, we find that most computational
aspects appear identically in the two-dimensional analogue. Appreciating the
advantages for visualisation we therefore present most of our observations in
this setting, adding 3D results mainly to confirm these findings. For structure
we will use the scenarios of Figure~\ref{fig:situation} as a baseline test
case, with minor modifications where required.

\subsection{Linear inversion: slip distribution}
\label{sec:linear}

\begin{figure}
\includegraphics[scale=.67,trim=0mm 0mm 10mm 8mm,clip]{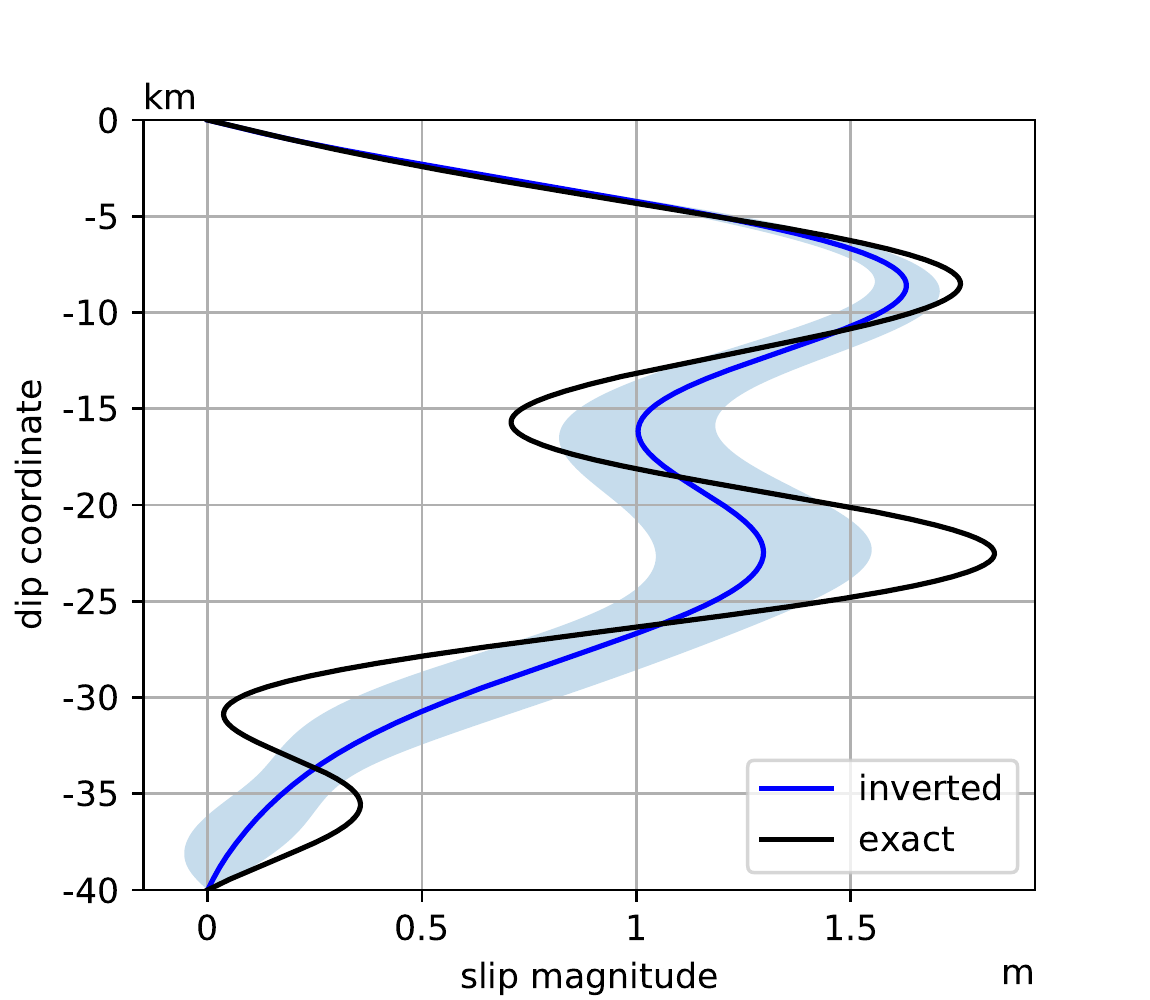}\hfill%
\includegraphics[scale=.67,trim=0mm 0mm 10mm 8mm,clip]{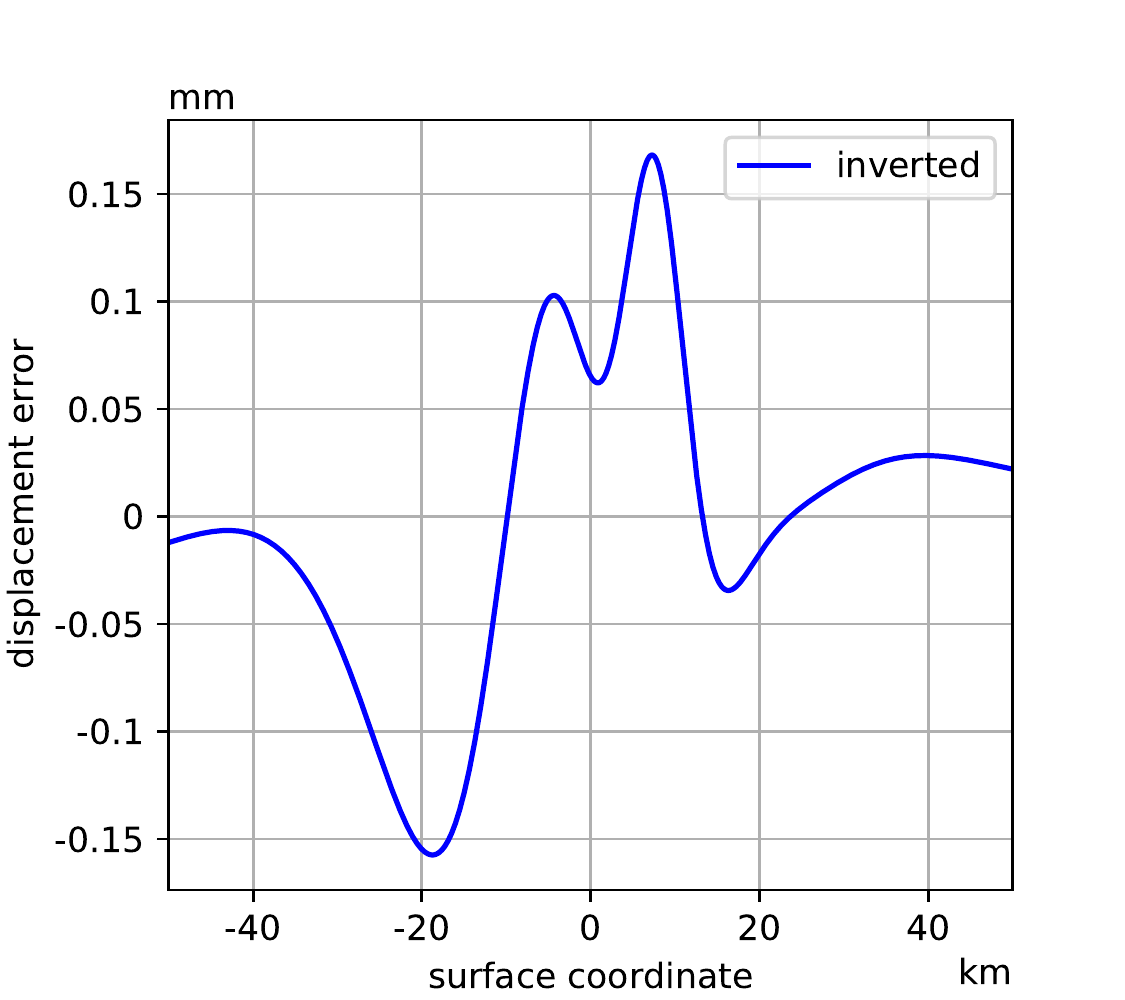}
\caption{Linear inversion of the slip distribution in the 2D non-rupturing
scenario of Figure~\ref{fig:situation} using Volterra's equation. Left: The
exact slip distribution $b$, the inverted slip distribution $b'(m,d)$, and the one
standard deviation, 68\% confidence interval $\pm\sqrt{h^T \Sigma_B'(m) h}$.
Right: The difference between the exact surface deformation and the deformation
that corresponds to the inverted slip distribution $b'(m,d)$ using Volterra's
equation as the forward model.}
\label{fig:exactlininv}
\vspace{1cm}
\includegraphics[scale=.67,trim=0mm 0mm 10mm 8mm,clip]{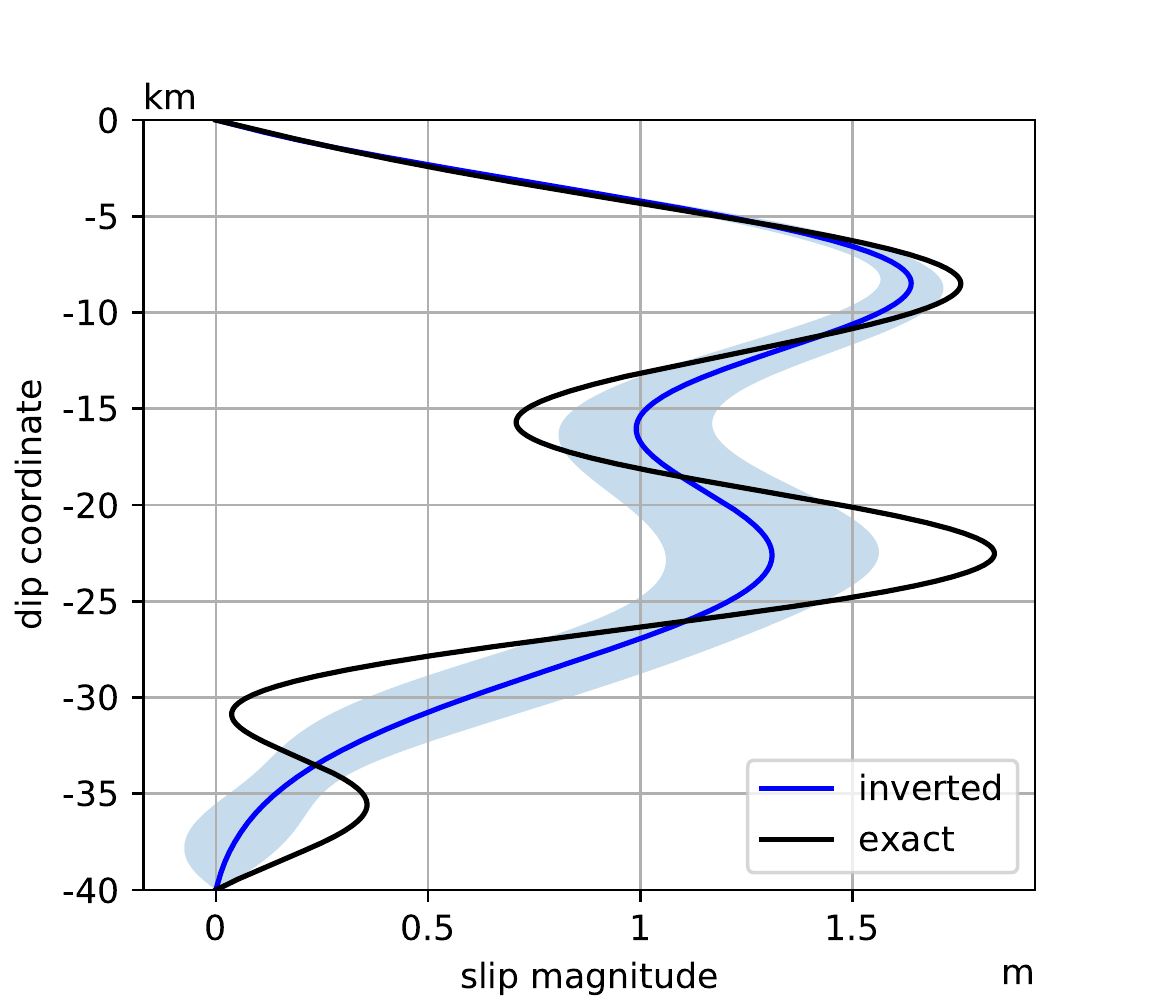}\hfill%
\includegraphics[scale=.67,trim=0mm 0mm 10mm 8mm,clip]{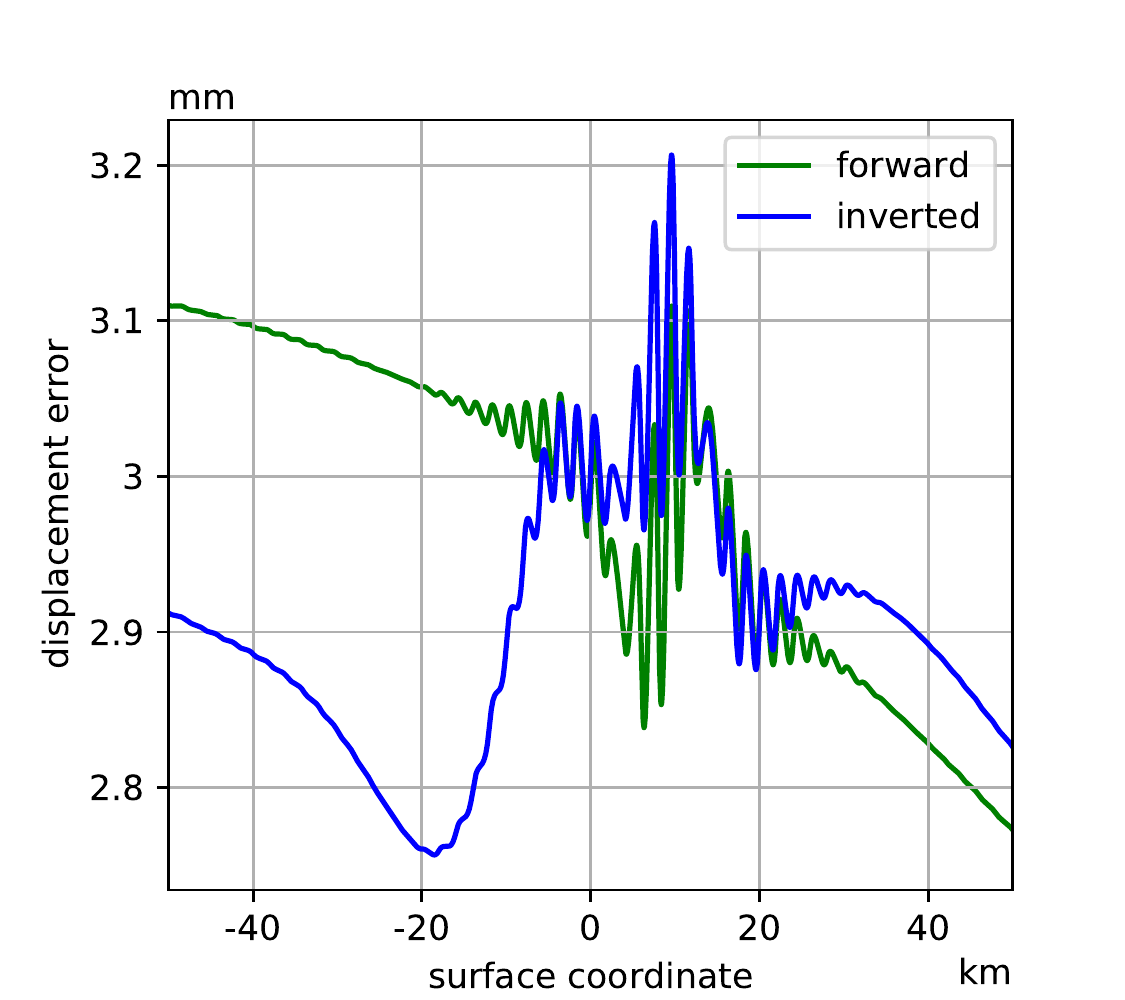}
\caption{Linear inversion of the slip distribution in the 2D non-rupturing
scenario of Figure~\ref{fig:situation} using the WSM forward model on a
$76\times 38$ element mesh, of which $50\times 25$ elements form the
search box. The graph layout is identical to that of
Figure~\ref{fig:exactlininv}, with the addition of the `forward' deformation
error that corresponds to the exact slip distribution, rather than the inverted
slip distribution.}
\label{fig:wsmlininv}
\end{figure}

We will study first the linear inversion process, in which we keep the fault
parameters $m$ equal to the exact values and invert the slip distribution $b$
only --- referring to the methodology of Section~\ref{sec:method} we set
$\tilde m = m$ in step~\ref{item:fwd}. A natural starting point is to establish
the best case solution by inverting using Volterra's equation, the forward
model that was used to synthesize the data. The results of this are shown in
Figure~\ref{fig:exactlininv}. While the inverted slip distribution does not
match the exact slip due to the smoothing effect of the prior distribution, we
observe in the left panel a reasonable fit that is in keeping with the
posterior variance. In the right panel we observe that the deformation error
stays well below the 1 mm measurement noise standard deviation.

Repeating this process with the WSM on a $76 \times 38$ element mesh we obtain
the result of Figure~\ref{fig:wsmlininv}, showing that the expected slip and
standard deviation are almost identical to those obtained using Volterra's
equation. Paradoxically, the corresponding deformation error in the right panel
(`inverted') is relatively large. It is noteworthy that the deformation error
is characterized by a significant offset (approximately 3 mm) with a relatively
small variation (approximately 0.2 mm). A similar offset can be seen in the
(`forward') error that the model produces with the exact slip as input, thus
representing the discretization error for this particular computational
setting. From this we can conclude that the offset does not result from the
inversion process, but is in fact a side effect of the discrete model.
Fortunately, by virtue of the differencing approach layed out in
Section~\ref{sec:synth}, the inversion is insensitive to offsets of this kind.
We observe that the errors are small relative to the offset, with a peak to
peak error range that is well below 1 mm, which explains the perceived paradox.

\begin{figure}
\includegraphics[scale=.67,trim=0mm 0mm 10mm 8mm,clip]{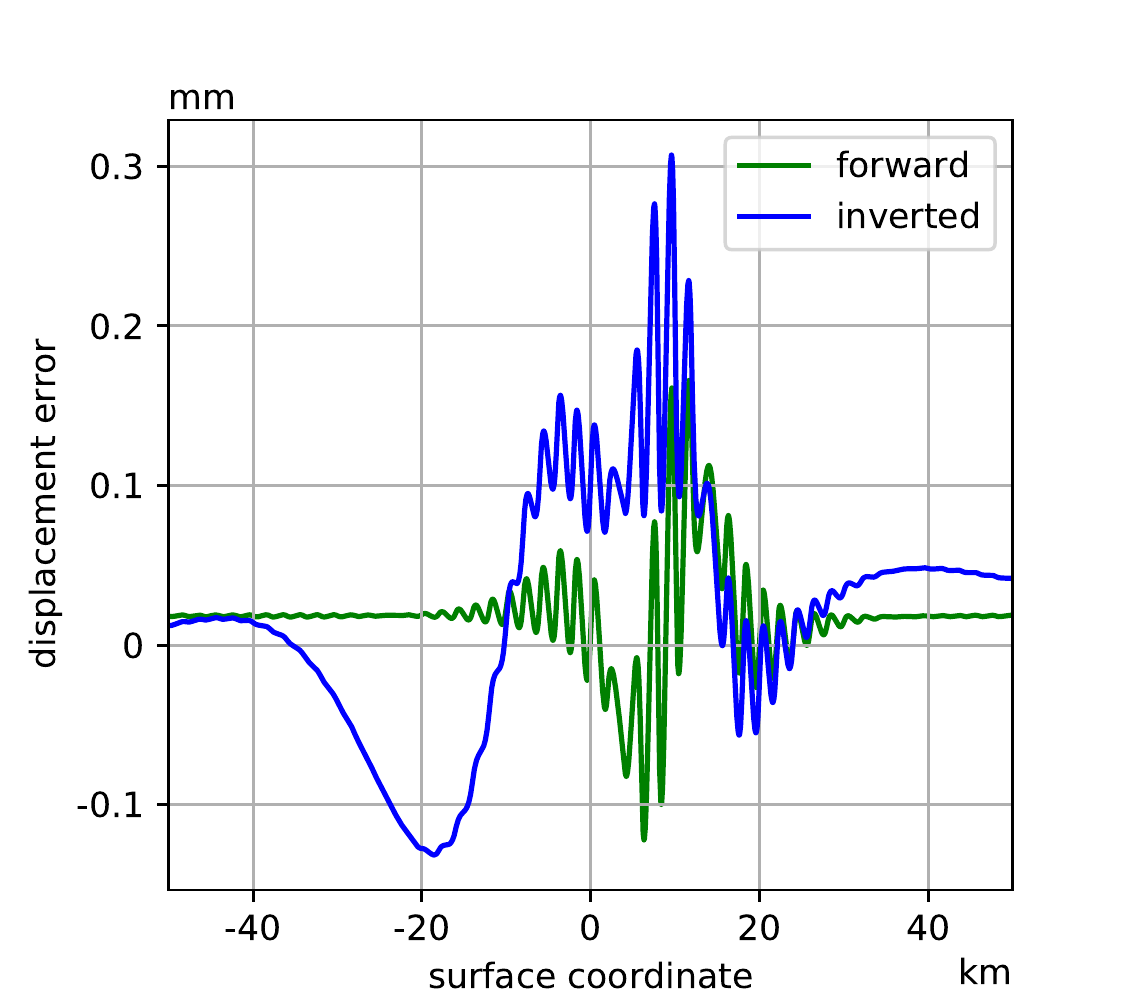}\hfill%
\includegraphics[scale=.67,trim=0mm 0mm 10mm 8mm,clip]{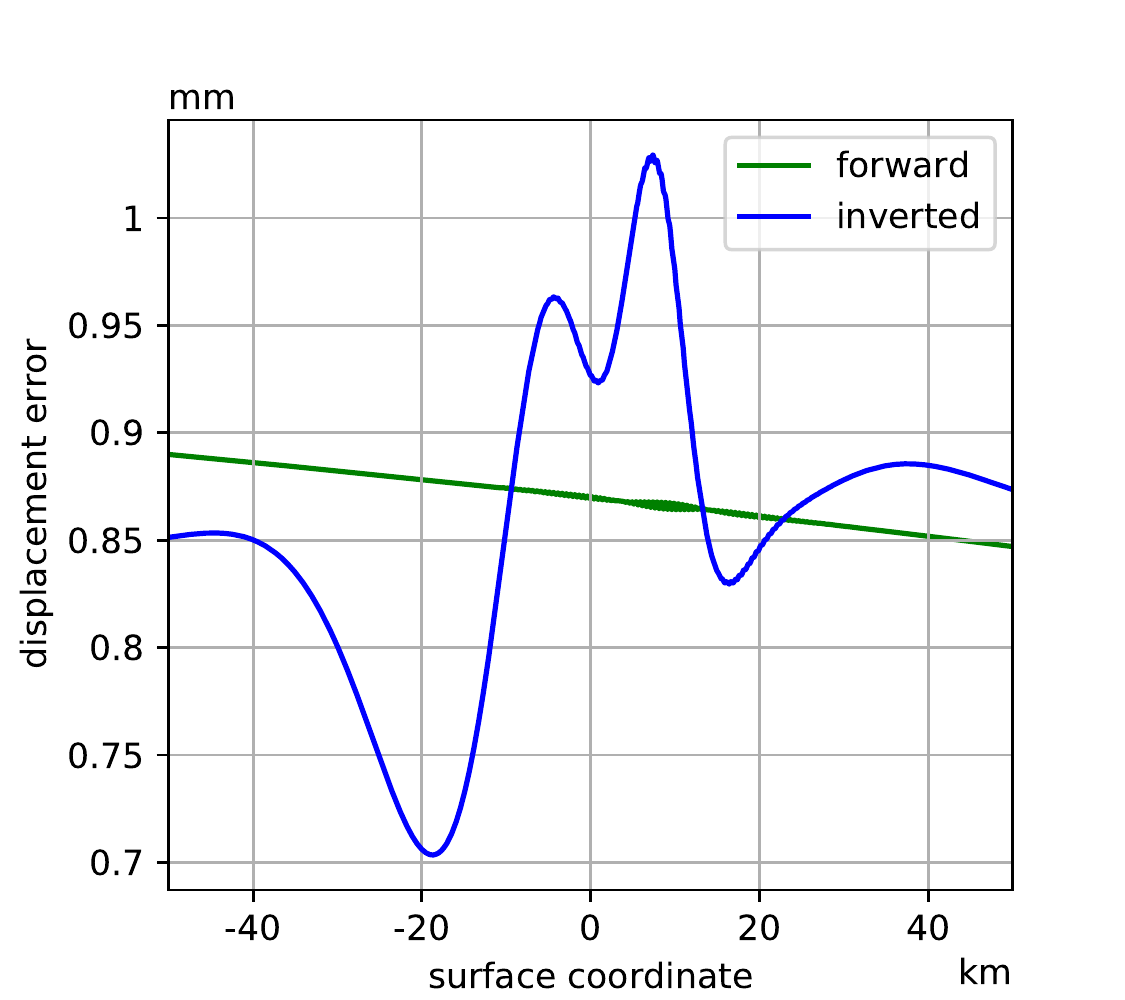}
\caption{Linear inversion of the slip distribution in the identical setting to
Figure~\ref{fig:wsmlininv} except for different finite element meshes.
Referencing Equation~\eqref{eqn:mesh}, the baseline mesh was constructed for
$n_{\rm box}=25$ and $n_{\rm inf}=38$, resulting in a $76\times 38$ element
mesh of which $50\times 25$ elements form the search box. Left: $n_{\rm
box}=25$ and $n_{\rm inf}=200$, keeping the search box resolution fixed while
adding elements to the far field. Right: $n_{\rm box}=100$ and $n_{\rm
inf}=150$, keeping the infinity-to-box ratio fixed while increasing resolution
by a factor 4.}
\label{fig:wsmmesh}
\vspace{1cm}
\includegraphics[scale=.67,trim=0mm 0mm 10mm 8mm,clip]{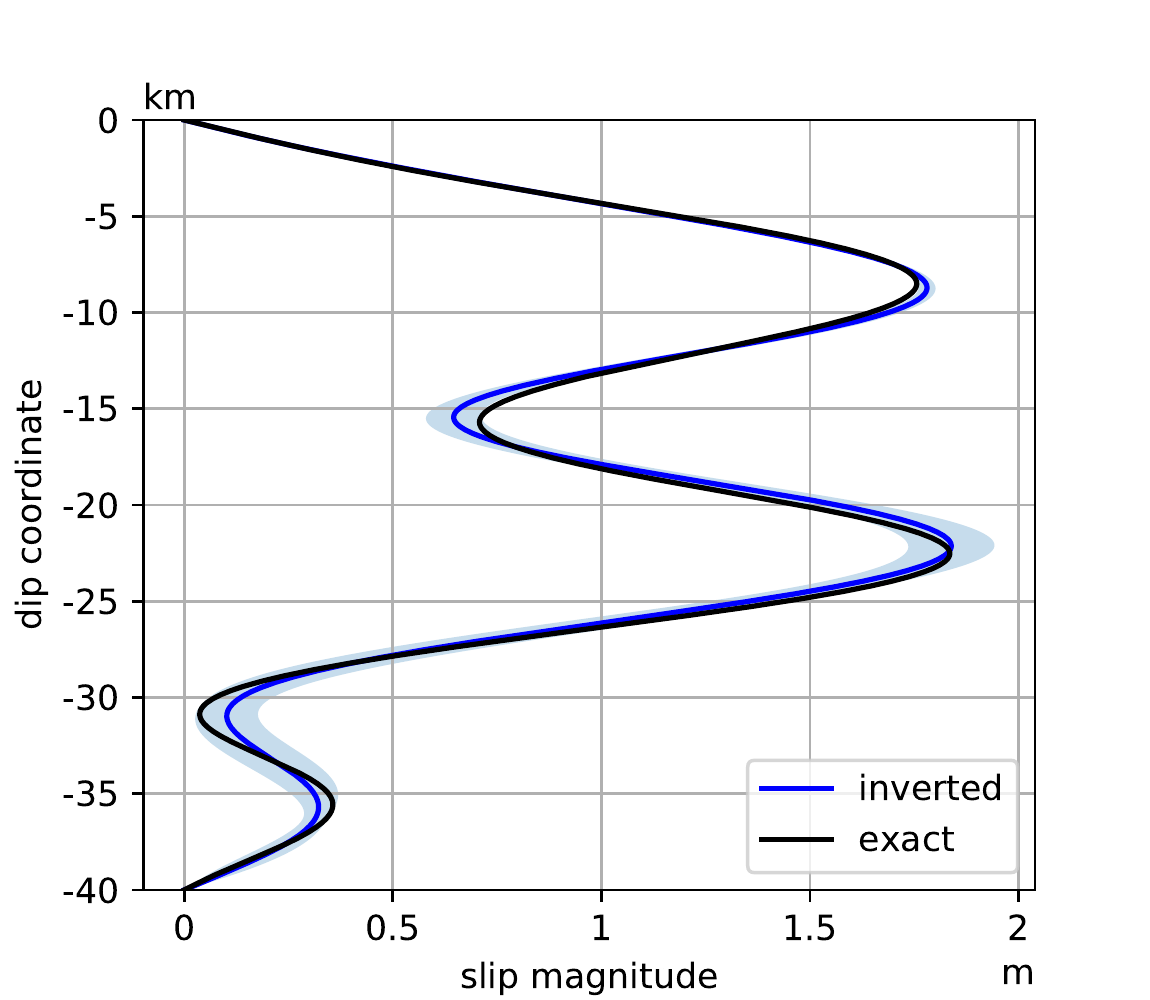}\hfill%
\includegraphics[scale=.67,trim=0mm 0mm 10mm 8mm,clip]{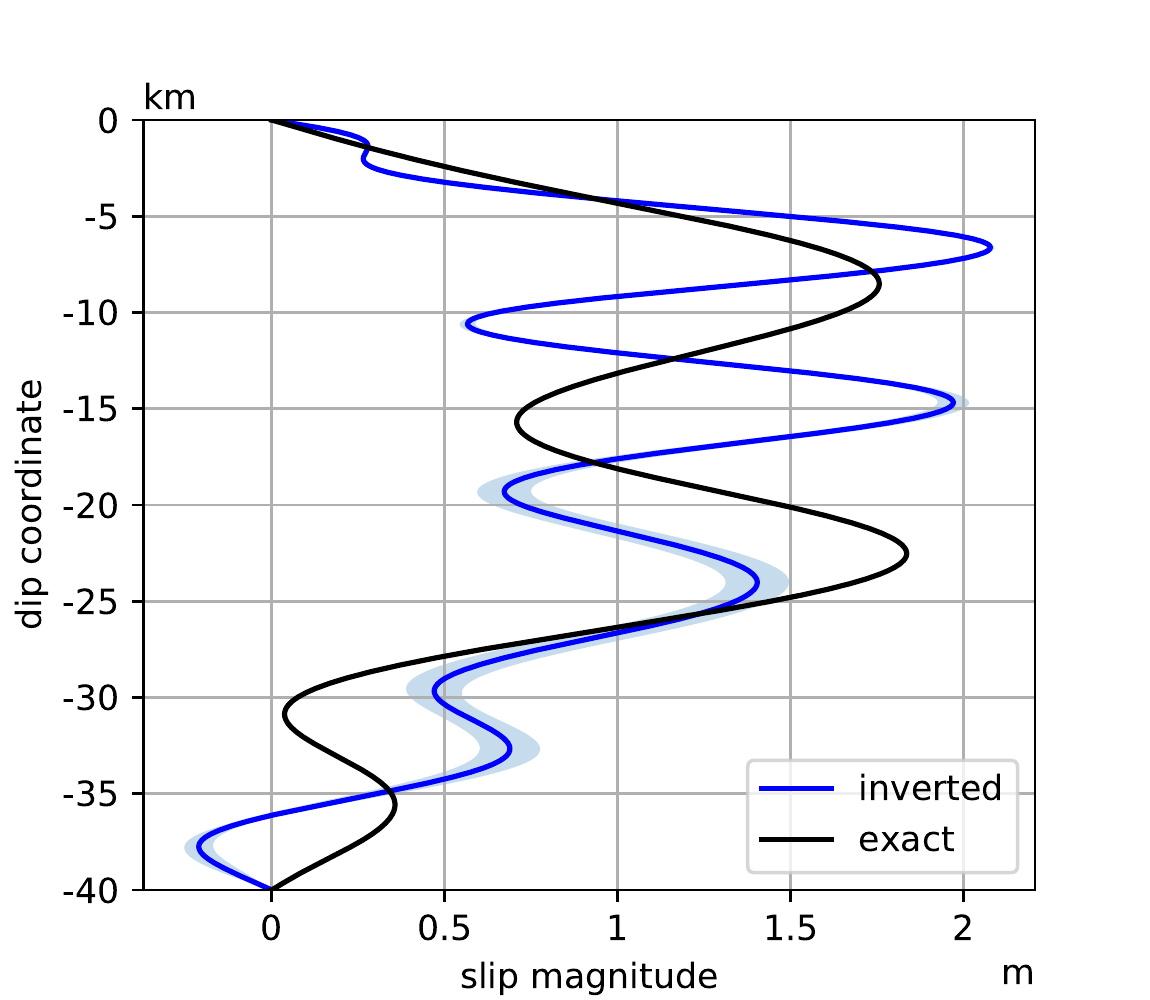}
\caption{Linear inversion of the slip distribution in the identical setting to
Figure~\ref{fig:wsmlininv} except for a 100$\times$ reduction of the
measurement noise to 0.01 mm. Left: Volterra-based inversion. Right: WSM-based
inversion on a $76\times 38$ element mesh, of which $50\times 25$ elements form
the search box. This is an illustration of the adverse effects on the inversion
when the noise level undercuts the discretization error.}
\label{fig:wsmlininvnoise}
\end{figure}

In addition to the 3 mm offset, the forward error curve of
Figure~\ref{fig:wsmlininv} shows a distinct spatial trend, dropping by 0.35 mm
over the length of the domain. Both aspects of the discretization error are
studied in Figure~\ref{fig:wsmmesh}, which shows two variations of the mesh
resolution. On the left we see the effect of increasing the resolution in the
far field while keeping that in the search box fixed. Comparing to
Figure~\ref{fig:wsmlininv}, we observe that both the offset and the trend are
greatly reduced, indicating that these phenomena are caused largely by the
treatment of the far field. At the same time the errors did not change
significantly relative to the offset, confirming that this far field-induced
error should not strongly affect the inversion. On the right we see how an
eight-fold uniform mesh refinement results in a stark reduction of the
discretization error, and in an inverted deformation error that closely
resembles that of the baseline result of Figure~\ref{fig:exactlininv} modulo
the remaining offset and trend.

Taken together, the results of Figure~\ref{fig:wsmmesh} uphold the result of
\cite{vanZwieten_2014} that the discretization error can be made arbitrarily
small via mesh refinement. However, as we have already seen, it is by no means
necessary to drive the error orders below the noise level of the deformation
measurements. To see what happens in the opposite direction,
Figure~\ref{fig:wsmlininvnoise} compares the Volterra and WSM-based inversion
at a noise level that is 100 times smaller while maintaining the mesh. While
Volterra's equation correctly tightens the error margin around the exact slip,
the WSM-based inversion deviates significantly from the exact slip due to the
dominant numerical error. At this noise level it takes an eightfold uniform
mesh refinement for the WSM-based inversion to return to being
indistinguishible from the Volterra based inversion, beyond which the inversion
is essentially mesh-independent. Based on these results, we consider that a
mesh at which the discretization error does not exceed half the standard
deviation of the measurement noise appears to strike a good balance between
accuracy and numerical efficiency.

There is one situation where we cannot control the discretization error through
mesh refinement, which is in the case of a rupturing fault. As the
approximation is inherently continuous, the error at the point of intersection
equals half the slip magnitude regardless of element size. Since this violates
the established rule that the discretization error may not exceed the
measurement noise, care must be taken to avoid the detrimental effects we
observed in the right panel of Figure~\ref{fig:wsmlininvnoise}. Arguably the
simplest way to achieve this is to discard measurements close to the rupture
and use only the remaining intermediate to far field data. Incidentally, the
masking out of data in the rupture zone is not uncommon in the context of SAR
interferometry, as local destruction tends to lead to decorrelation of the radar signal.
We hypothesize, therefore, that no valuable data need be discarded in practice.

\begin{figure}
\includegraphics[scale=.67,trim=0mm 0mm 10mm 8mm,clip]{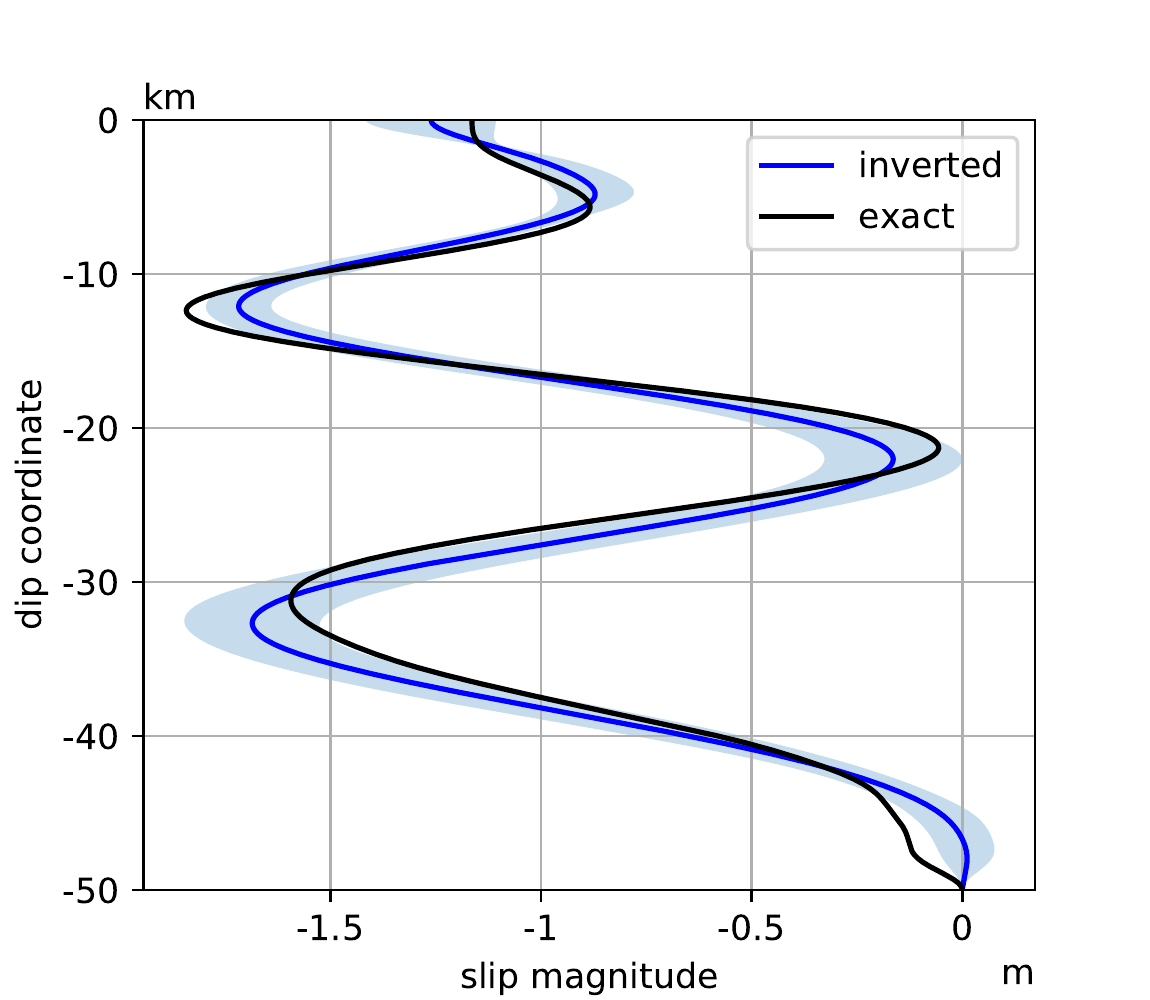}\hfill%
\includegraphics[scale=.67,trim=0mm 0mm 10mm 8mm,clip]{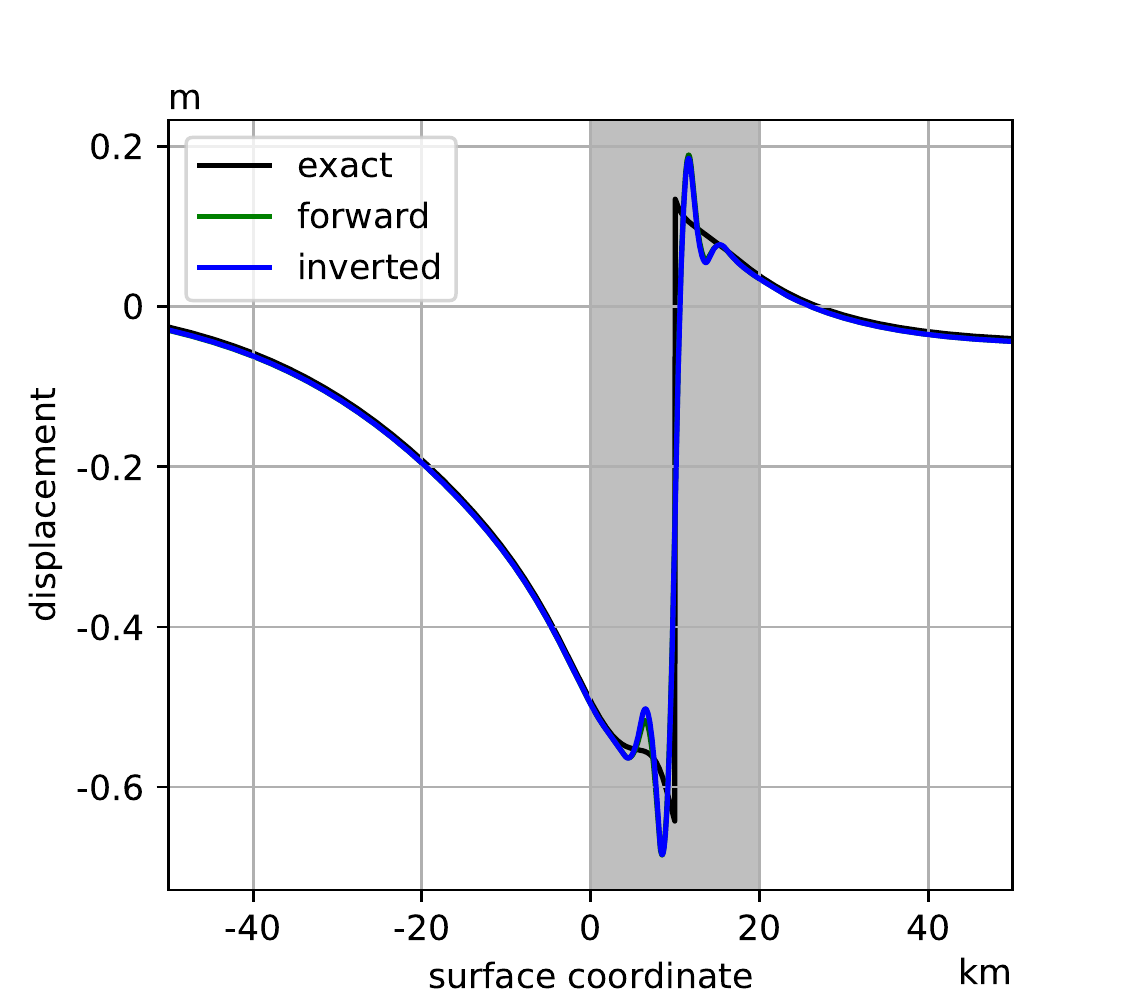}
\caption{Linear inversion of the slip distribution in a 2D rupturing scenario.
The graph layout is similar to that of Figure~\ref{fig:exactlininv}, with the
difference that the right panel shows the absolute displacements rather than
the displacement errors with the black `exact' curve representing the
synthesized displacement field. The gray band in the the right panel
corresponds to the area that was masked out in order for the locally
meter-scale errors not to affect the inversion.}
\label{fig:rupturelininv}
\vspace{1cm}
\includegraphics[scale=.67,trim=0mm 0mm 10mm 8mm,clip]{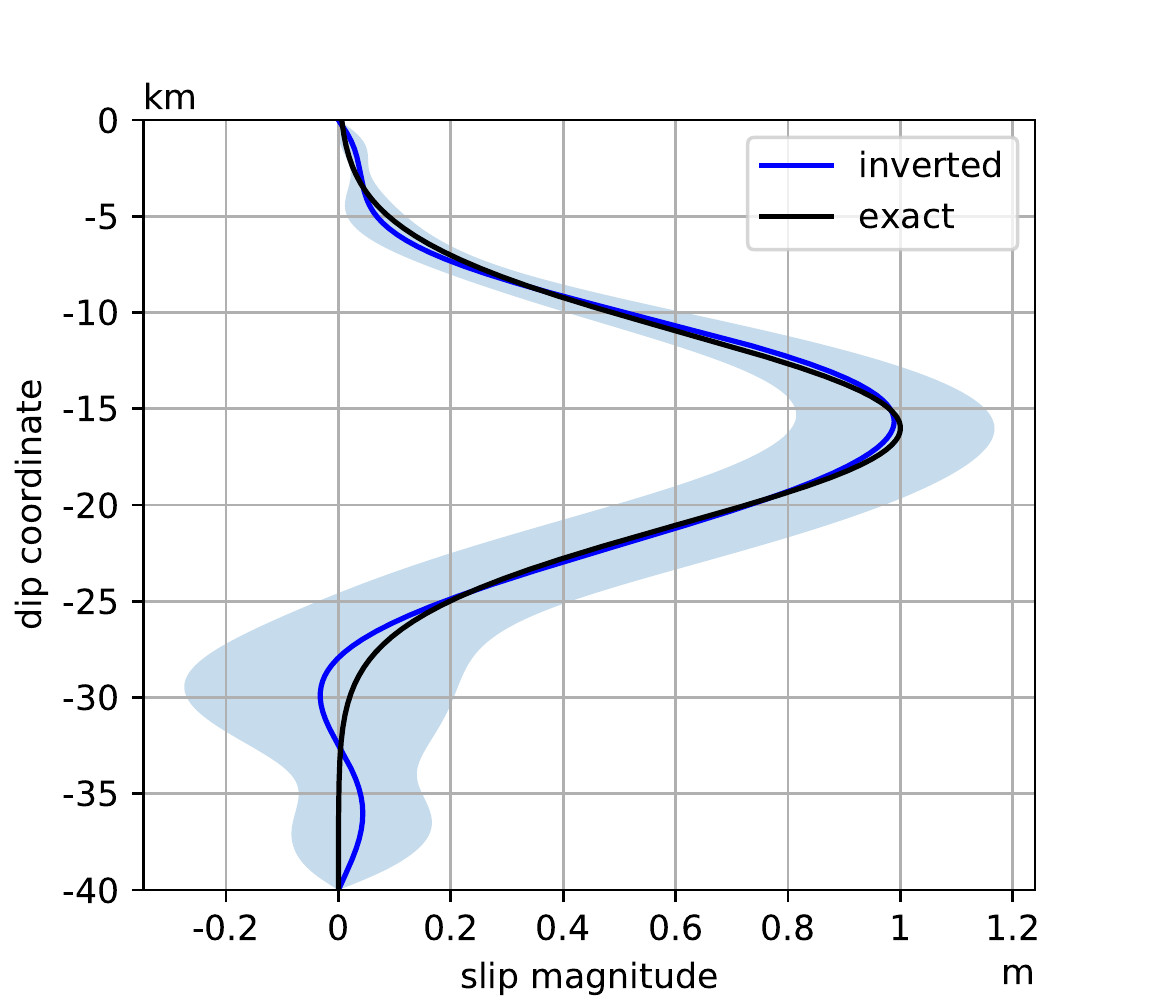}\hfill%
\includegraphics[scale=.67,trim=0mm 0mm 10mm 8mm,clip]{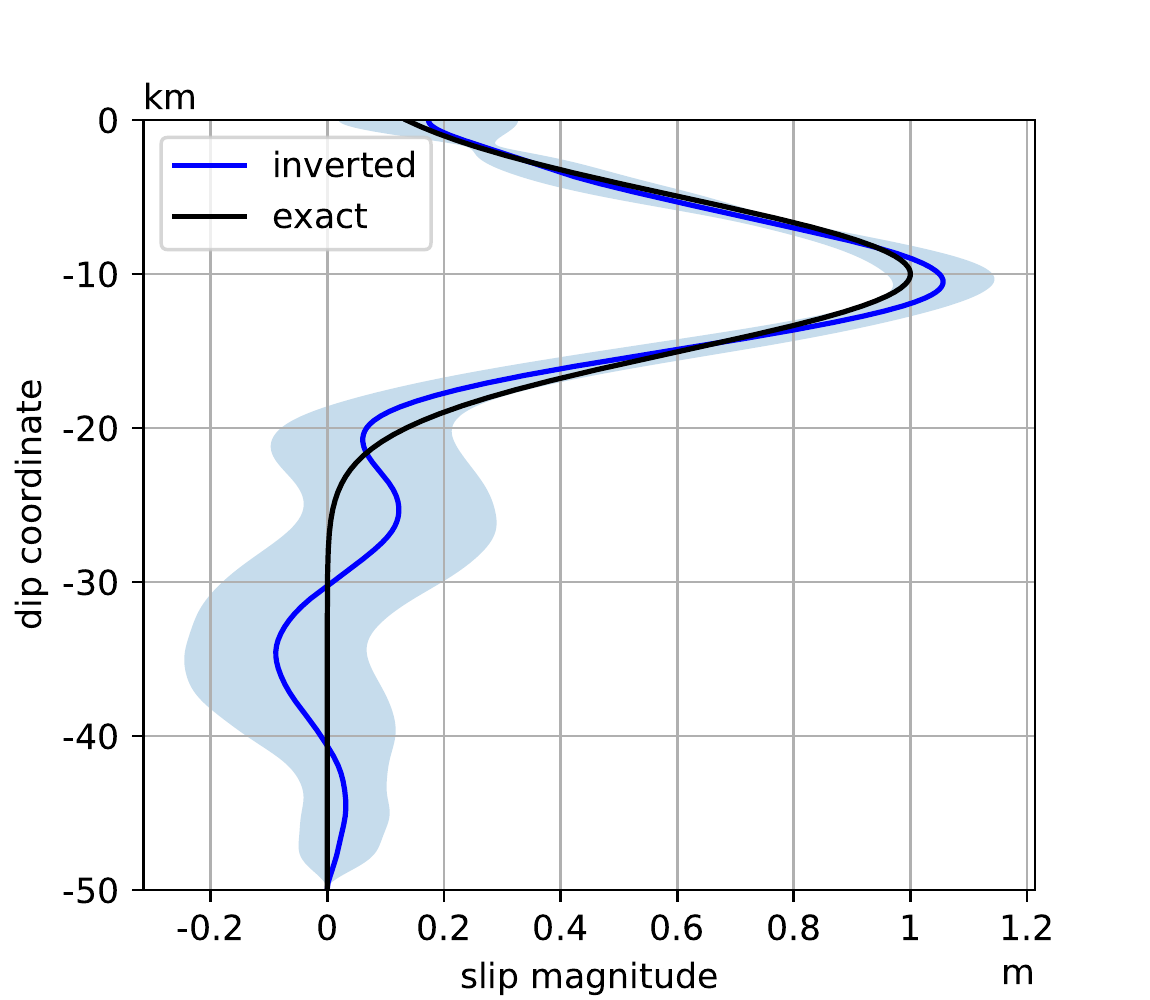}
\caption{Linear inversion of a manually constructed Gaussian slip distribution
in a 2D scenario, solved on a $76\times 38$ element mesh. Left: a non-rupturing
scenario with the slip centered at 40\% dip. Right: a rupturing scenario with
the slip centered at 20\% dip and a masked area of 20 km.}
\label{fig:localslip}
\end{figure}

An example of a rupturing fault can be seen in Figure~\ref{fig:rupturelininv}.
The right panel shows the absolute displacement (rather than the displacement
error) in which we observe continuous oscillations at the 10 km position where
the exact displacement exhibits a discontinuity. The oscillations decay
rapidly, reaching sub-millimeter scale amplitudes at a 10 km distance from the
surface rupture. The left panel shows the inversion result based on the
deformation data outside of this $\pm 10$ km interval that is marked gray in
the right panel. The result accurately recovers the exact slip distribution,
and is virtually indistinguishable from the Volterra-based inversion (not
displayed) subject to the same data mask, confirming that data masking is a
suitable strategy to deal with the continuous representation of discontinuities
in the WSM.

So far we have drawn a slip distribution from the prior distribution, the same
that is subsequently used in the inversion procedure to reconstruct the slip
from measurements. Since it is difficult in practice to accurately capture
prior knowledge in terms of a distribution, it is relevant to study the
robustness of the procedure to slip distributions not being elements of our
discrete space $\mathcal B$. Examples of this can be seen in
Figure~\ref{fig:localslip}, which shows two Gaussian slip distributions, one
rupturing, the other non-rupturing. Though neither is a member of $\mathcal B$,
both distributions are recovered with
reasonable accuracy, confirming that the
methodology has at least some lenience to inadequacies in the choice of the
prior. This also confirms our premise that the size of the fault plane need not
be an independent parameter if we have reasonable upper bounds, as the areas of
zero slip are captured accurately.

\begin{figure}
\includegraphics[scale=.67,trim=25mm 0mm 32mm 8mm]{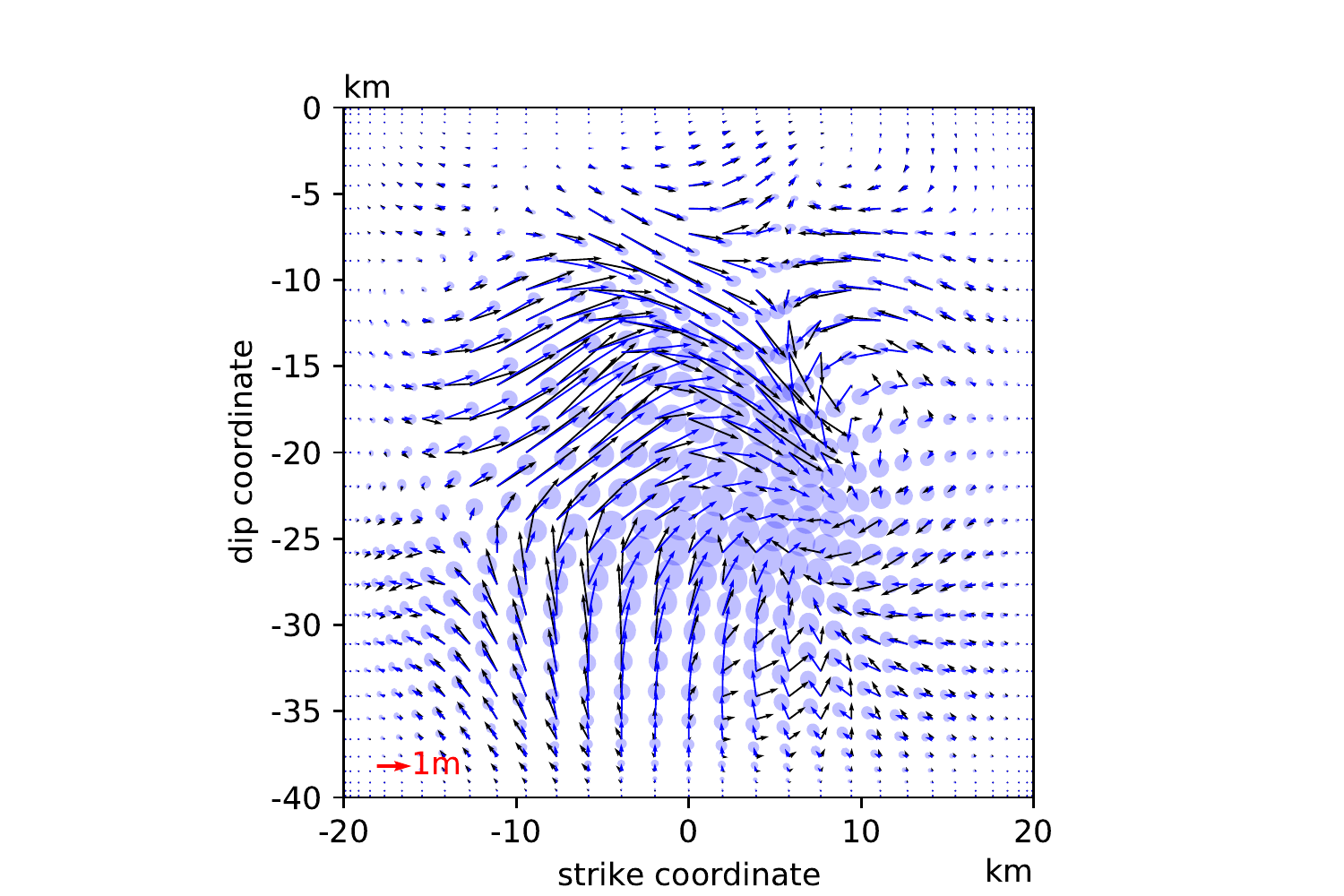}\hfill%
\includegraphics[scale=.67,trim=26mm 0mm 17mm 8mm,clip]{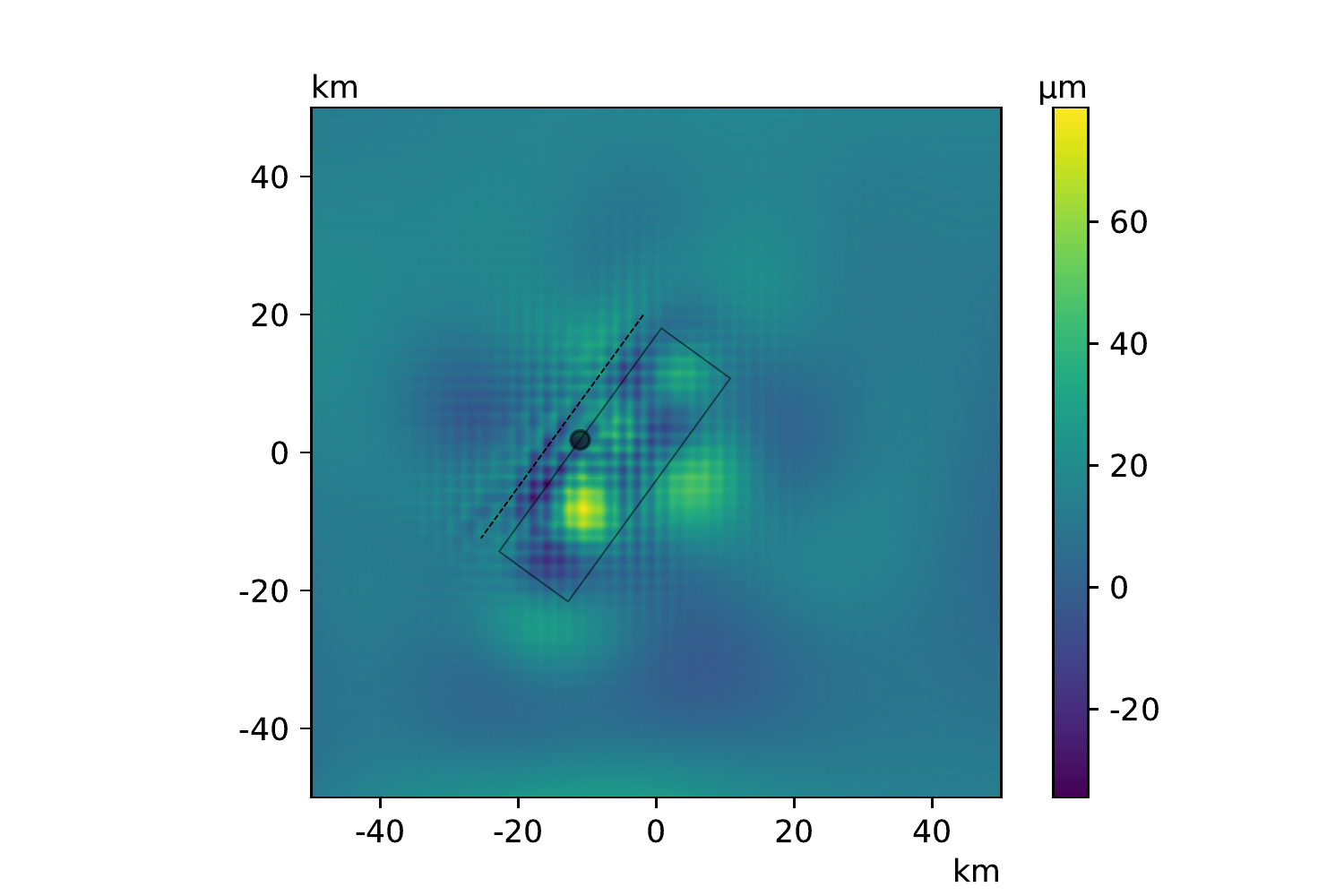}
\caption{Linear inversion of the slip distribution in a 3D non-rupturing
scenario using the WSM forward model on a $76\times 76\times 38$ element mesh,
of which $50\times 50\times 25$ elements form the search box.
Left: The exact slip distribution $b$ as black quivers, the inverted slip
distribution $b'(m,d)$ as blue quivers, and the local one standard deviation or
47\% confidence region resulting from the $2\times 2$ posterior covariance
matrix $h^T \Sigma_B'(m) h$ as blue ellipses, centered at the expected value
and at matching scale. Right: the vector norm of the deformation gradient
error. The solid rectangle shows the outline of the fault plane.}
\label{fig:lininv3d}
\end{figure}

Finally we turn to the 3D scenario (right panel) of Figure~\ref{fig:situation}.
Employing the identical methodology, Figure~\ref{fig:lininv3d} shows the slip
distribution and the corresponding error in deformation gradient, in which we
recognize similar patterns to those we observed in the direct 2D equivalent of
Figure~\ref{fig:wsmlininv}. Taking into account the local standard deviation,
the inverted and exact slip distributions are in good agreement. We also
confirmed that the result is indistinguishable from that obtained via
Volterra's equation.
The error is largest in the vicinity of the fault, but stays well clear of the
1 mm noise level. Interestingly, the error offset that we observed in the 2D
results is much less pronounced in the 3D situation.

\subsection{Nonlinear inversion: fault parameters}
\label{sec:nonlinear}

We proceed by studying the nonlinear inversion of the fault parameters.
Referring again to the methodology of Section~\ref{sec:method}, in
step~\ref{item:fwd} we now evaluate the posterior expected value for the fault
parameters $m$ using the Metropolis-Hastings MCMC process. This process samples
the posterior distribution through repeated evaluation of $f_d(m)$ as defined
in Equation~\eqref{eqn:mcmcfunc}, being proportional to the posterior
probability density function $f_{M|D}$. As this entails evaluation of the
expected value $b'(m,d)$, all aspects of the linear inversion process as
explored in the previous section remain in effect.

\begin{figure}
\includegraphics[scale=.67]{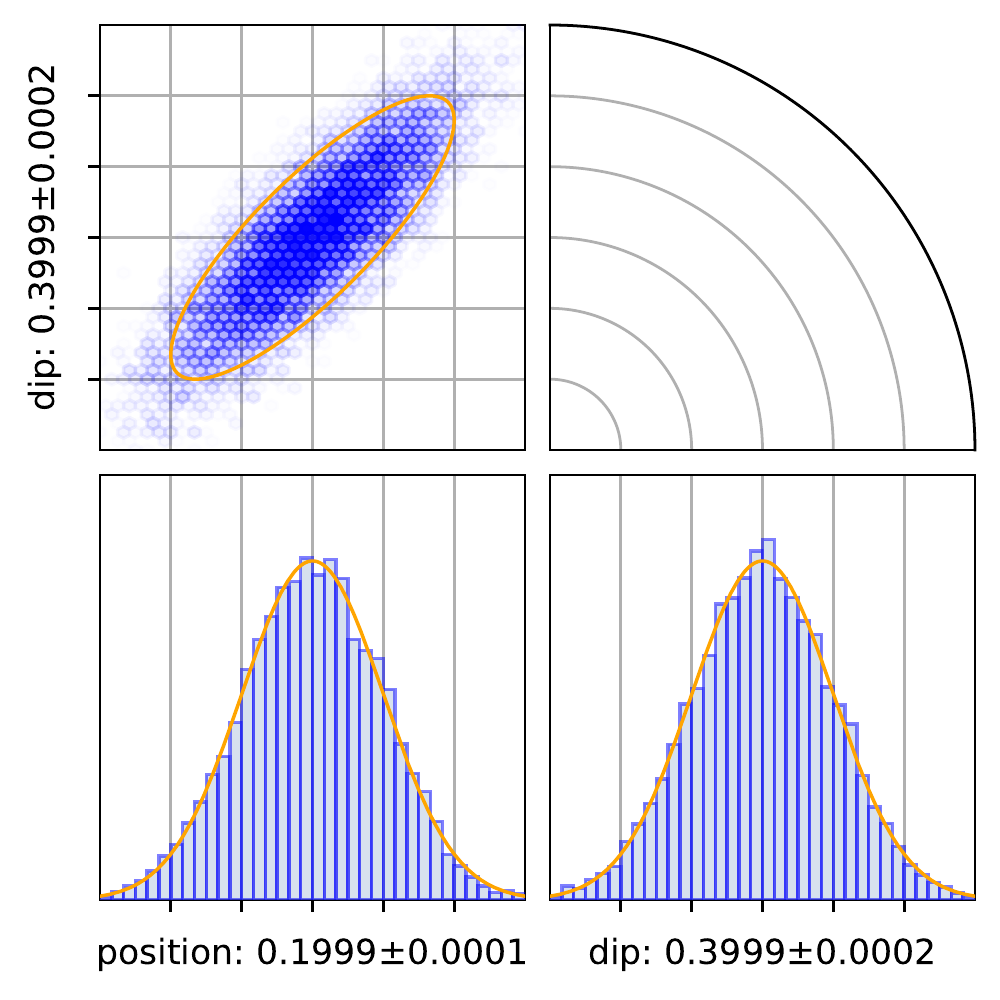}\hfill%
\includegraphics[scale=.67]{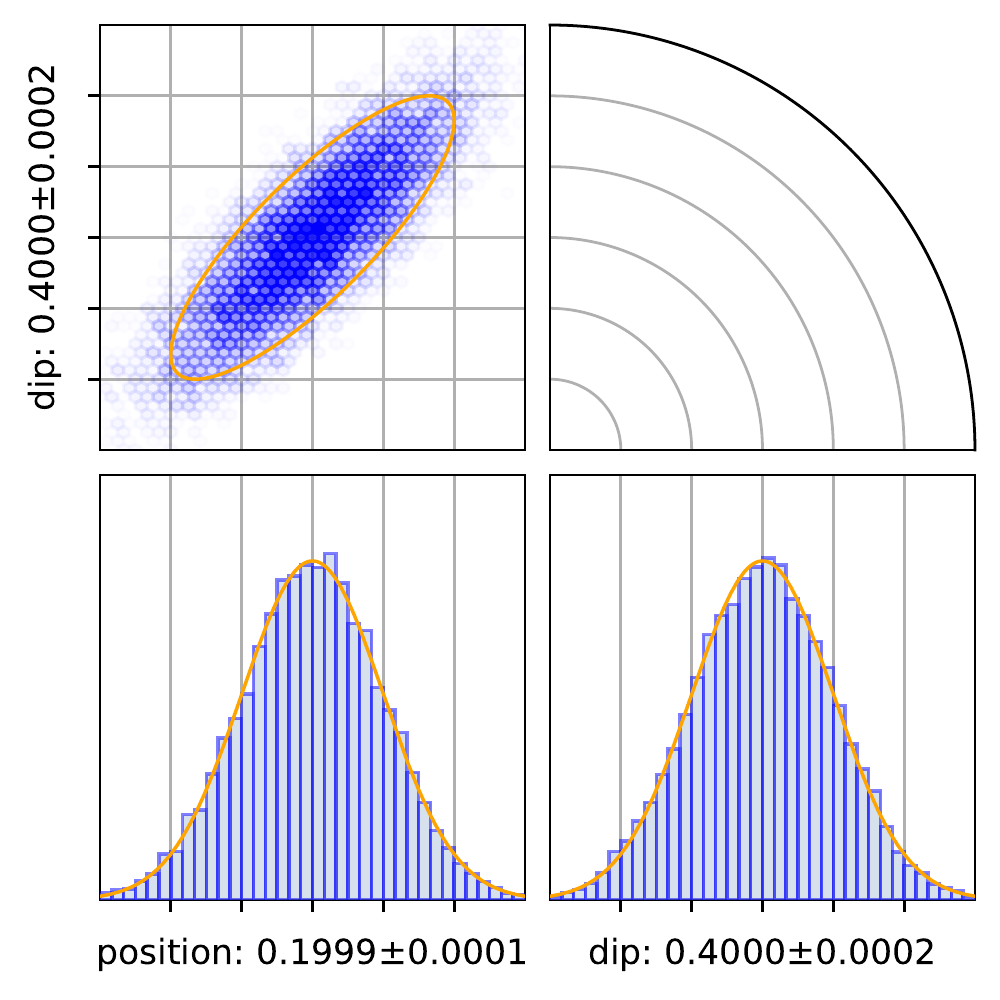} \\[-2mm]
{\small\rule{0pt}{0pt} \hspace{12mm} position [$L$] \hspace{12mm} dip angle [$\pi$] \hspace{28mm} position [$L$] \hspace{12mm} dip angle [$\pi$]}
\caption{Binned results of a 50.000-sample MCMC process for $M|D$ using
Volterra's equation in the 2D non-rupturing scenario of
Figure~\ref{fig:situation}, which has fault parameters position=$0.2L$ and dip
angle=$0.4\pi$. Left: reference results from Volterra's equation. Right:
results from the WSM on a $48\times 24$ element mesh, of which $32\times 16$
elements form the search box. The bottom row shows the marginalized
distributions for fault position (left) and the dip angle (right), with axis
labels showing mean value $\pm$ standard deviation or, equivalently, the 68\%
confidence interval. The orange overlay shows the corresponding normal
distribution. Grid lines are spaced at one standard deviation. The top row
shows the cross correlation of x coordinate and dip angle, with the orange
overlay showing the bivariate normal distribution at two standard deviations
or, equivalently, the 91\% confidence region.}
\label{fig:nonlin-nonrupt}
\vspace{1cm}
\includegraphics[scale=.67,trim=8mm 2mm 13mm 10mm,clip]{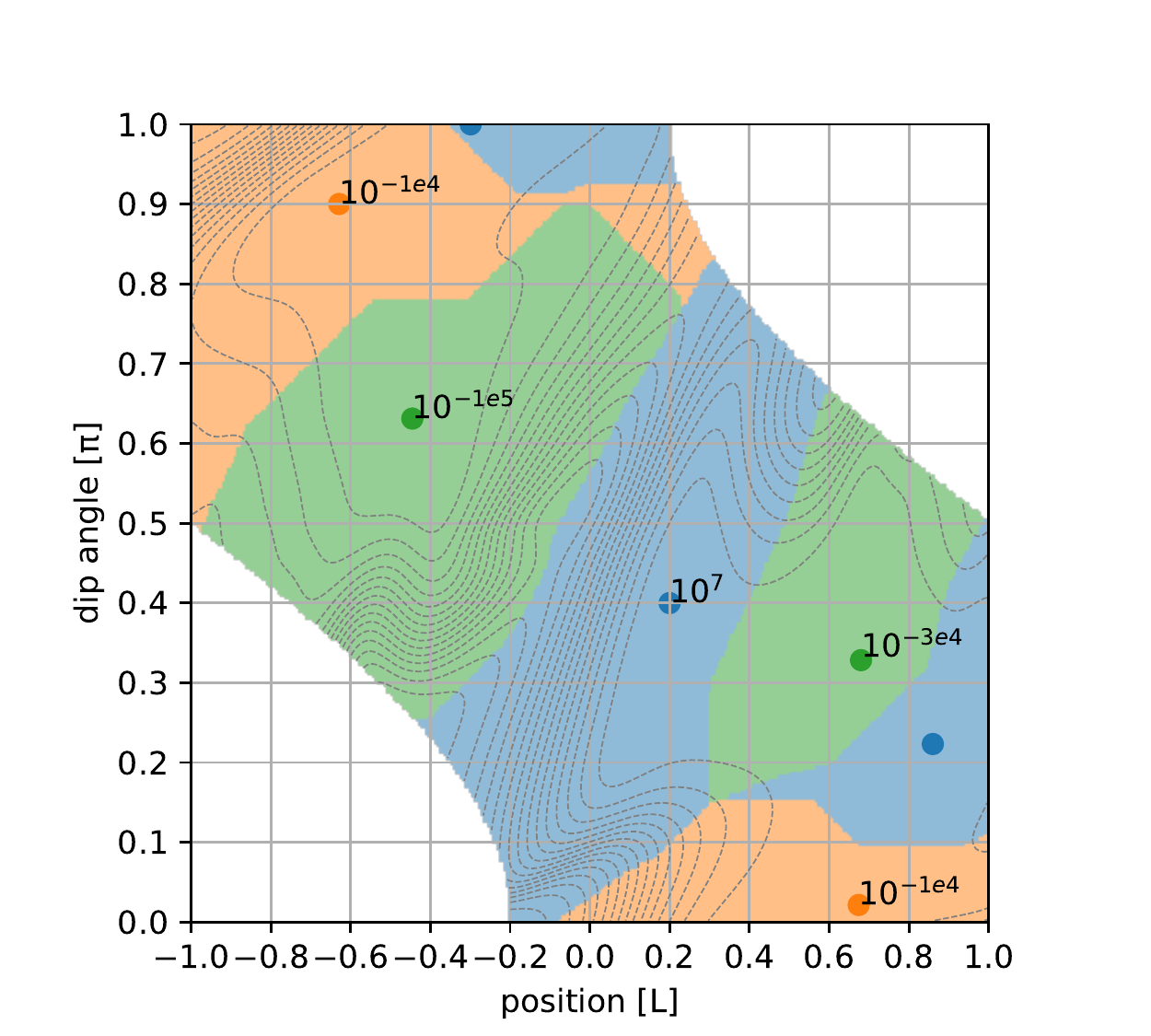}\hfill%
\includegraphics[scale=.67,trim=8mm 2mm 13mm 10mm,clip]{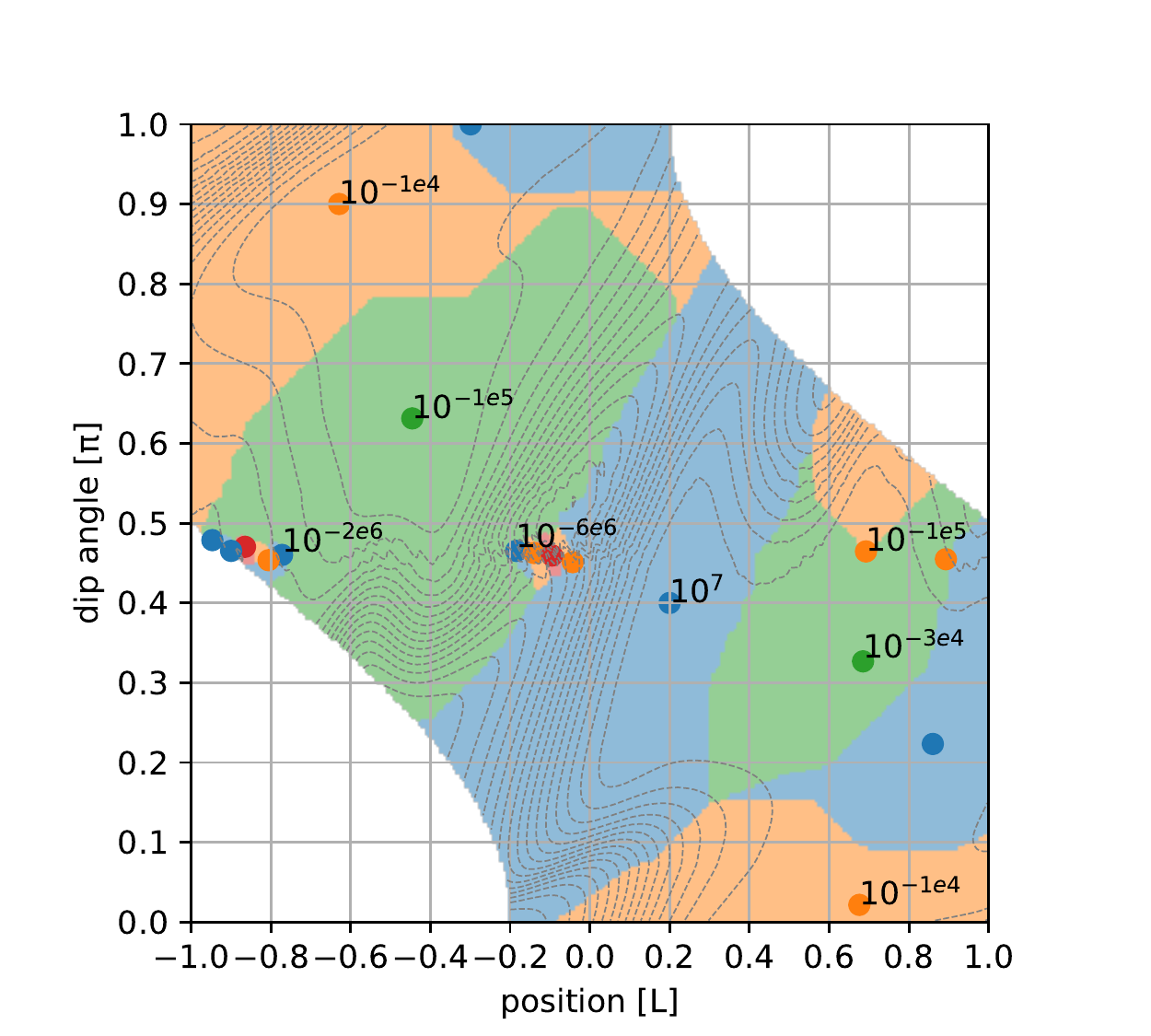}
\caption{Contour lines and local maxima of $f_{M|D}$ for the same 2D
non-rupturing scenario as that of Figure~\ref{fig:nonlin-nonrupt}. The coloured
areas represent (inverse) watersheds of the probability landscape, that is, the
collection of points from which an uphill gradient method converges to the
associated local maximum. Left: reference result from Volterra's equation.
Right: results from the WSM on a $48\times24$ element mesh.}
\label{fig:landscape-nonrupt}
\end{figure}

Returning to the non-rupturing 2D baseline scenario (left panel) of
Figure~\ref{fig:situation}, Figure~\ref{fig:nonlin-nonrupt} shows the results
of a MCMC process comparing Volterra's equation to the WSM. We use the same
relatively coarse mesh that we used for Figure~\ref{fig:wsmlininv} to see if
there are adverse effects in pushing against the boundary of the discretization
error. Both distributions are seen to capture the fault parameters correctly,
pinpointing the exact values to a high degree of accuracy both in terms of a
nearly exact expected value and of the equally narrow confidence interval.
Furthermore, both distributions are in excellent agreement to each other,
demonstrating the robustness of the method with regard to discretization
errors.

We started the MCMC random walk from the global maximum, obtained via the
Nelder-Mead uphill simplex method, that in turn was started from where we know
the exact solution to be. Knowledge of an exact solution is a luxury that is
not available in any practical application, which means a global search
algorithm will typically be required to prepare for the final gradient ascent.
A relevant question, therefore, is whether the WSM has a disturbing effect in
this regard. To explore this, Figure~\ref{fig:landscape-nonrupt} compares the
posterior probability density function obtained through direct evaluation of
Volterra's equation against that obtained via the WSM, identifying local maxima
as well as the associated watersheds. While the WSM introduces some spurious
local maxima, the global maximum as well as its watershed appear identical,
suggesting that there is no difference with regard to global optimization
strategies.

\begin{figure}
\includegraphics[scale=.67]{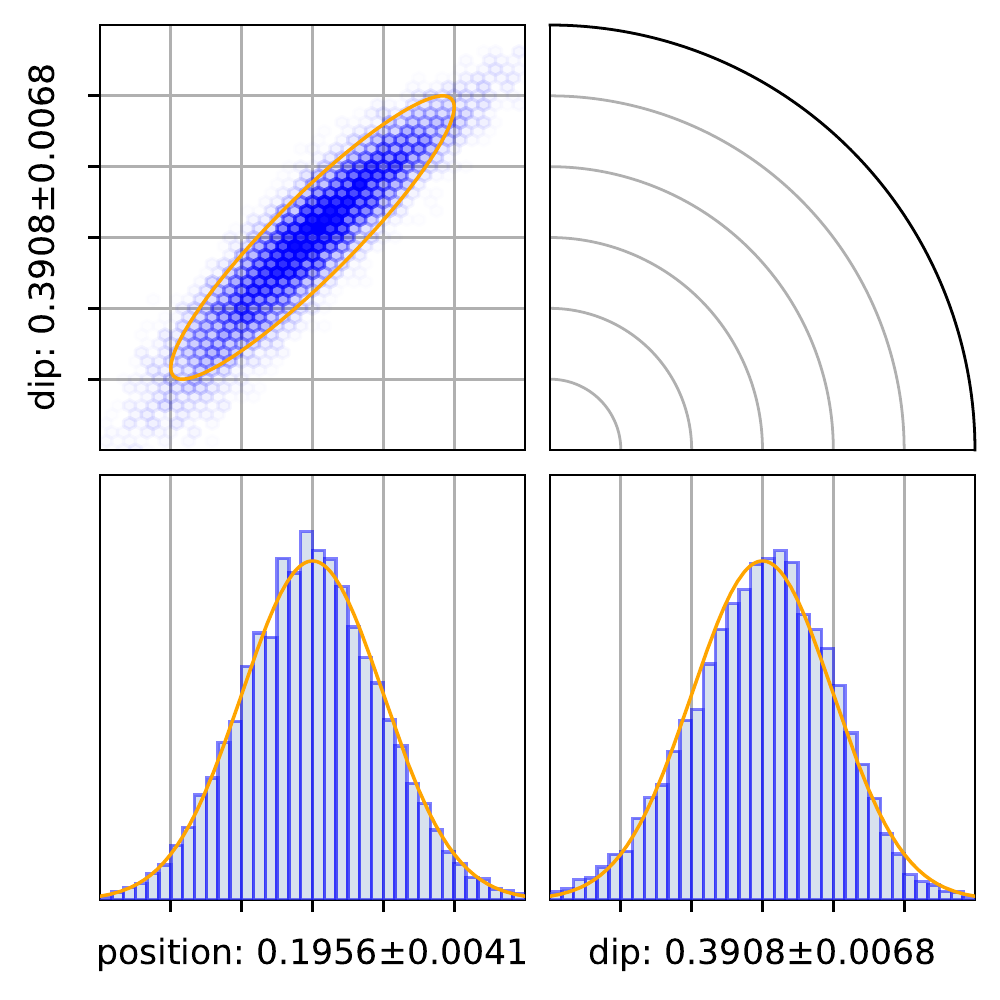}\hfill%
\includegraphics[scale=.67]{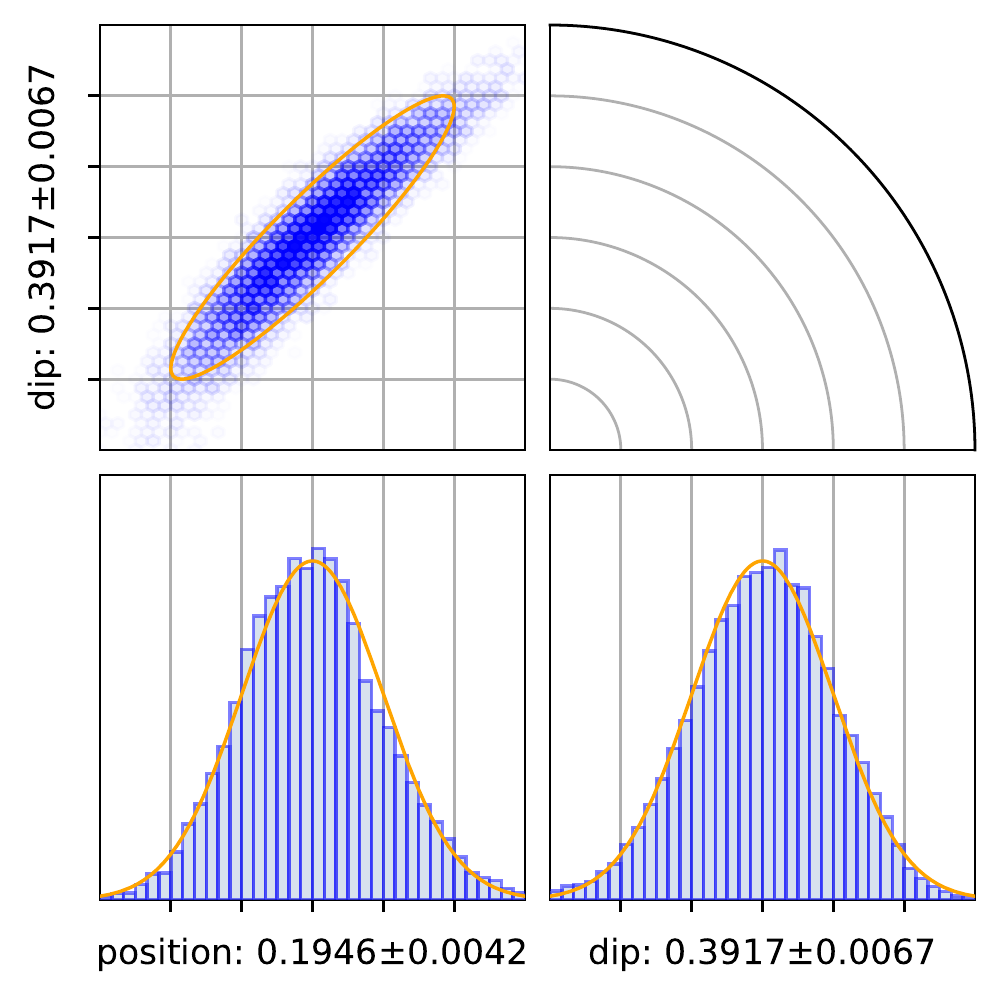} \\[-2mm]
{\small\rule{0pt}{0pt} \hspace{12mm} position [$L$] \hspace{12mm} dip angle [$\pi$] \hspace{28mm} position [$L$] \hspace{12mm} dip angle [$\pi$]}
\caption{Binned results of a 50.000-sample MCMC process for $M|D$ using the WSM
forward model in a 2D rupturing scenario with the same parameters as in
Figure~\ref{fig:nonlin-nonrupt}: position=$0.2L$ and dip
angle=$0.4\pi$. A radius $0.3L$ data mask is applied centered
at x=$0.2L$. Left: reference results from Volterra's equation. Right:
results from the WSM on a $96\times 48$ element mesh, of which $64\times 32$
elements form the search box.}
\label{fig:nonlin-rupt}
\vspace{1cm}
\includegraphics[scale=.67,trim=8mm 2mm 13mm 10mm,clip]{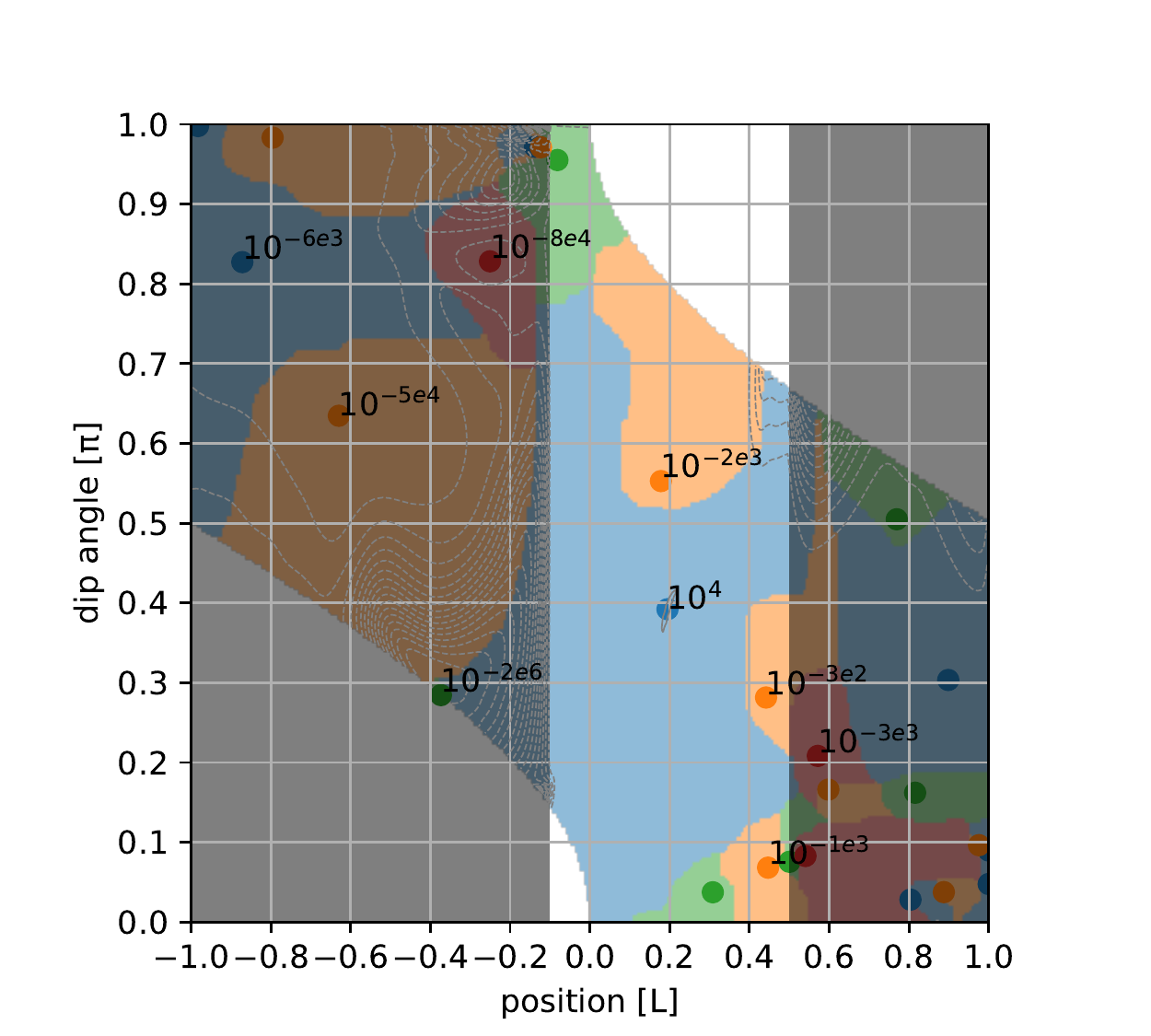}\hfill%
\includegraphics[scale=.67,trim=8mm 2mm 13mm 10mm,clip]{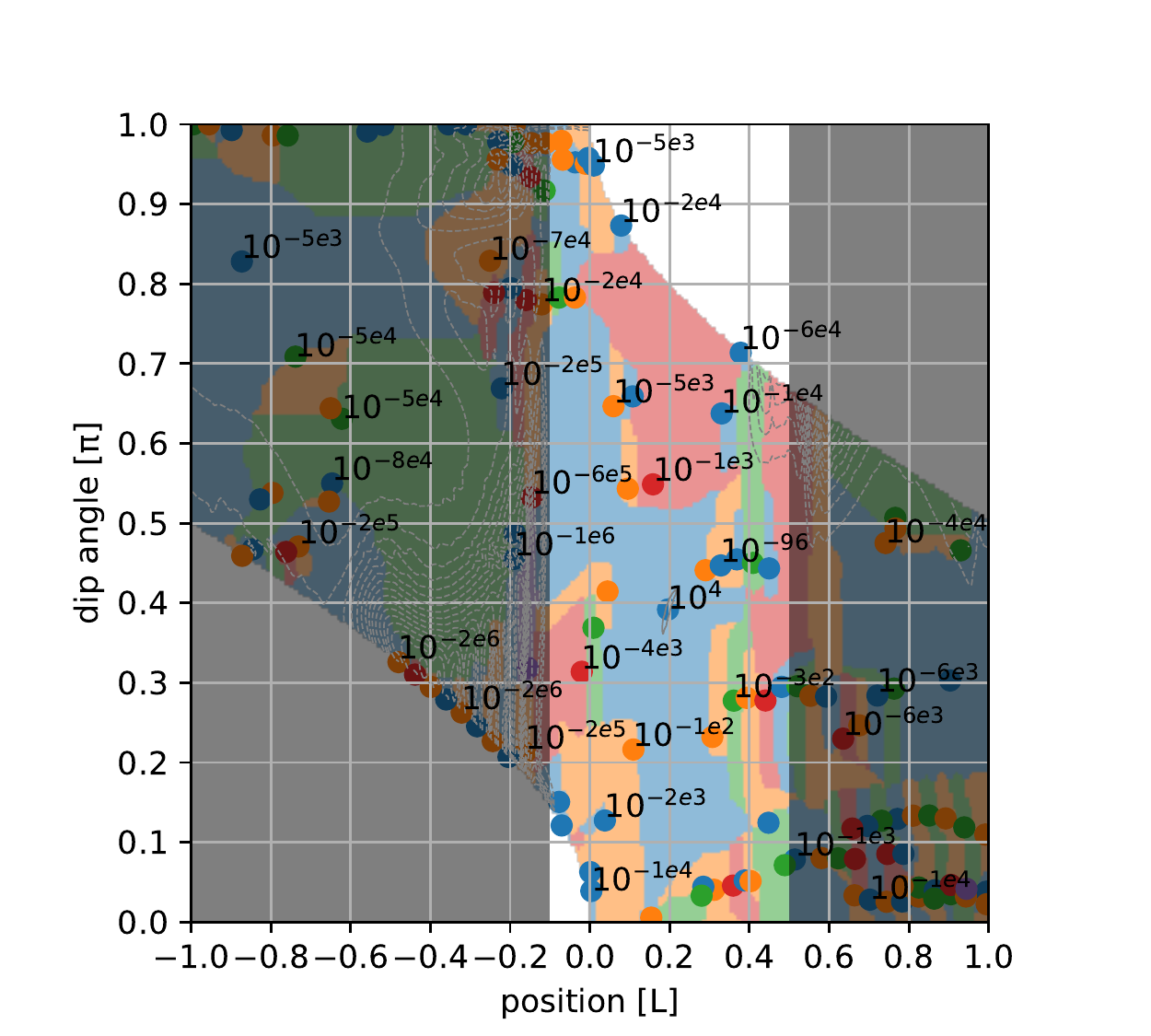}
\caption{Contour lines, local maxima and watersheds of $f_{M|D}$ for the same
2D non-rupturing scenario as that of Figure~\ref{fig:nonlin-rupt}, with an
overlay delineating the masked region. Left: reference result from Volterra's
equation. Right: results from the WSM on a $96\times48$ element mesh.}
\label{fig:landscape-rupt}
\end{figure}

For the rupturing scenario we mask out an area of 30 km centered at the point
of rupture. While it may seem contradictory to mask out the rupture zone while
simultaneously inverting for the rupture coordinate, this process should be
understood in the context of having prior information in the form of in situ
observations. Even though the precise trajectory of the rupture may not be
known to sufficient accuracy, it may well be sufficient to define a masking
zone. In fact, though not employed here, we are at liberty to modify the prior
distribution to have it reflect this knowledge as well. Selecting the uniformly
refined grid of Figure~\ref{fig:rupturelininv}, the 30 km zone is 50\% wider
than the observed minimum in order to not artificially limit mobility of the
rupture coordinate, but rather give it some freedom to find its optimum within
the confines of the broader mask.

Figure~\ref{fig:nonlin-rupt} shows the side by side results of Volterra's equation
and the WSM. Note that the mask was applied to Volterra's equation as well for
sake of comparison, even though the method does not require it. The
distributions in the rupturing scenario are less precise as a result of data
masking, but are otherwise in excellent agreement.

Figure~\ref{fig:landscape-rupt} again explores the posterior probability
density function, where this time we see a very large qualitative difference
between Volterra's equation and the WSM. While both show a clear delineation at
$(0.2\pm0.3)L$, corresponding to the applied data mask, the WSM produces many
more local maxima, clustering in particular at the crossover points and at
shallow dip angles. The global maximum still has a fairly large associated
watershed, however, suggesting that the multitude of local maxima is not
necessarily problematic in a global optimization context. Note also that the
global optimization algorithm needs only consider positions inside the masked
region --- the non-shaded region in the figure --- as ruptures outside of the
mask are in violation of its premise.

\begin{figure}
\includegraphics[scale=.67]{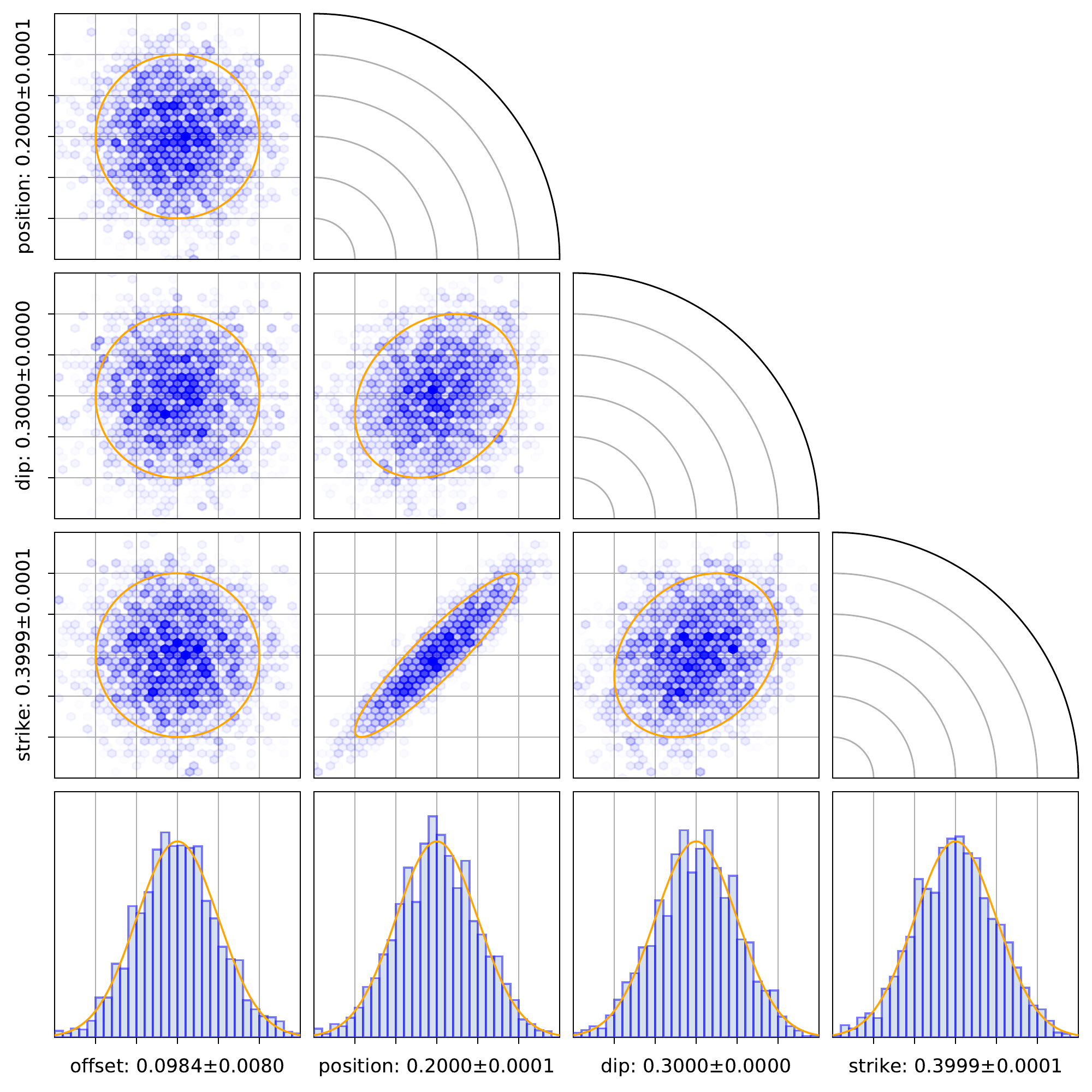} \\[-2mm]
{\small\rule{0pt}{0pt} \hspace{15mm} offset [$L$] \hspace{15mm} position [$L$] \hspace{12mm} strike angle [$\pi$] \hspace{10mm} dip angle [$\pi$]}
\caption{Binned results of a 10.000-sample MCMC process for $M|D$ corresponding
to the 3D non-rupturing scenario of Figure~\ref{fig:situation}, which has fault
parameters offset=$0.1L$, position=$0.2L$, strike angle=$0.3\pi$ and dip
angle=$0.4\pi$. The bottom row shows from left to right the marginalized
distributions for fault offset, position, strike and dip, with axis labels
showing mean value $\pm$ standard deviation or, equivalently, the 68\%
confidence interval. The orange overlay shows the corresponding normal
distribution. Grid lines are spaced at one standard deviation. The remaining
rows show from top to bottom the fault parameters y, strike and dip, thus
covering all cross correlations. The orange overlay shows the bivariate normal
distribution at two standard deviations or, equivalently, the 91\% confidence
region.}
\label{fig:mcmc}
\end{figure}

We conclude again with the 3D scenario (right panel) of
Figure~\ref{fig:situation}. Using the same mesh as was used for the slip
inversion of Figure~\ref{fig:lininv3d}, Figure~\ref{fig:mcmc} shows the
posterior distribution of the 3D fault parameters. One can observe that the
expected values accurately match the parameters that were used to generate the
synthetic data. The position along strike has a markedly larger variance than
that perpendicular to it, which matches the expectations laid out in
Section~\ref{sec:faultparams} relating to ambiguities with the slip
distribution. Though we cannot feasibly run the MCMC process using Volterra's
equation to verify the correctness of this result, we consider the foregoing to
be sufficient support to present this as a demonstration of the WSM driving a
realistic, nonlinear inversion of a 3D fault plane.

\section{Conclusions}
\label{sec:conclusions}

In this paper we performed for the first time a full inversion of fault plane
parameters and fault slip distribution using the Weakly-enforced Slip Method
(WSM) that was developed explicitly for this purpose. By restricting the domain
to a homogeneous halfspace we were able to synthesize deformation data, as well
as compare the WSM-based inversion against a reference result obtained from the
exact solution. This allowed us to study in detail the effect that the
discretization error has on different aspects of the inversion process. To
provide our study with a mathematical framework we placed the inverse problem
in a Bayesian setting, in which a prior probability distribution is refined
though observations into a posterior probability that quantities the likelihood
of various faulting mechanisms.

For linear inversions, the WSM was found to be competitive with Volterra's
equation in terms of accuracy, showing excellent agreement already at coarse
meshes (5 elements per 10 km for the situation considered) in case of
non-rupturing faults. As a practical rule of thumb for the minimum required
mesh density, we demonstrated empirically that the discretization errors must
not exceed the standard deviation of the measurement noise in order to avoid
large numerical errors. Conversely, increasing the mesh density beyond this
point contributes little to the accuracy of the inversion.

A WSM-based inversion of rupturing faults requires additional measures to
account for the local smearing out of the discontinuity, but it is argued that
similar measures are often required in practice regardless. When a simple data
mask is applied to disregard data points in the vicinity of the rupture, the
WSM and exact method again show excellent agreement, albeit at a finer mesh
that is required to localize the discrization error to the rupture zone. Local
to the rupture the error is observed to decay exponentially, at a rate that is
inversely proportional to the element size of the computational mesh. It is
expected that this relation holds as well in the case of local, rather than
uniform, refinements.

While the WSM was originally analyzed on a finite domain with exact boundary
conditions~\cite{vanZwieten_2013}, real world applications cannot rely on the
availability of such data. Instead of introducing artificial boundaries, we
opted for a finite-to-infinite mapping for the treatment of the far field in
order not to introduce assumptions that might limit the validity of our
results. Though this treatment introduces a fairly significant error, we
observe that the relative displacement error of two nearby points remains
dictated by the local element size. This circumstance fits remarkably well with
the fact that satellite-based InSAR observations are inherently relative, which
means that treatment of the data must be insensitive to global offsets. While
successful in this regard, meshing the far field is arguably an expensive
solution, increasing the number of degrees of freedom by a factor 3.38 in the
test cases considered. Further study in this direction is therefore warranted.

For non-linear inversions, any global search algorithm followed by a
gradient-based optimization is shown to perform equally well for the WSM as it
does for the exact forward method, as demonstrated by a full comparison of the
posterior probability density. Furthermore, we demonstrated that the iterates
of the Nelder-Mead uphill simplex method can be reused in a linear least
squares projection to provide a high quality Gaussian proposal distribution for
a subsequent exploration of the posterior probability using the
Metropolis-Hastings Markov Chain Monte Carlo method.

In an observation that is unrelated to the WSM we have remarked that certain
parameters that are customarily added to the space of fault parameters, such as
the fault plane dimensions, can instead be captured at lesser cost by the slip
distribution. In situations where an ambiguous relationship remains between a
fault parameter and the slip distribution, this translates to a large variance
of the posterior distribution, as demonstrated by the along-strike offset of
the fault plane in the 3D scenario.

In conclusion, we believe that the present work convincingly demonstrates the
utility of the WSM in real world applications, combining the power and
flexibility of finite element analysis with a highly efficient reuse of
computational effort. It also provides a practical framework by which such
studies can be performed. While the current experiments have been restricted to
homogeneous halfspaces for reasons of verification, none of these restrictions
were required by the methodology as it is presented here; nor do we have reason
to believe that our findings are limited to these conditions.

\bibliographystyle{plain}
\bibliography{literature}

\end{document}